\documentclass[twocolumn,numberedappendix, appendixfloats,apjl]{open_journal}

\usepackage{newtxtext,newtxmath}
\usepackage[T1]{fontenc}

\DeclareRobustCommand{\VAN}[3]{#2}
\let\VANthebibliography\thebibliography
\def\thebibliography{\DeclareRobustCommand{\VAN}[3]{##3}\VANthebibliography}

\usepackage{graphicx}	
\usepackage{amsmath}	
\usepackage{gensymb}
\usepackage[dvipsnames]{xcolor}
\usepackage{booktabs}
\usepackage{multirow}
\usepackage{enumitem}
\usepackage{bigdelim}
\usepackage[normalem]{ulem}
\setlist[itemize]{align=left, labelindent=\parindent, itemindent=0pt,leftmargin=*}
\setlist[itemize]{topsep=4pt}  
\setlist[itemize]{parsep=0pt, partopsep=0pt}  
\setlist[enumerate]{align=left, labelindent=\parindent, itemindent=0pt,leftmargin=*}
\usepackage[colorlinks=true
  ,urlcolor=blue
  ,anchorcolor=blue
  ,citecolor=blue
  ,filecolor=blue
  ,linkcolor=blue
  ,menucolor=blue
  ,linktocpage=true
  ,pdfproducer=medialab
  ,pdfa=true
]{hyperref}
\definecolor{orcidlogocol}{HTML}{A6CE39}
\newcommand{\OrcidID}[1]{ \href[urlcolor = red]{https://orcid.org/#1}{\textcolor{lightgray}{\faOrcid}}}
\newcommand{\OrcidIDName}[2]{\href{https://orcid.org/#1}{\color{blue}{#2}}}

\defcitealias{Arico2024}{A24}
\defcitealias{Mead2020}{M20}

\newcommand*{\refc}[1]{%
  \begingroup
    \hypersetup{
      linkcolor=Mahogany,
      linkbordercolor=Mahogany,
    }%
    \ref{#1}%
  \endgroup
}
\newcommand{\murl}[1]{%
  {\hypersetup{urlcolor=magenta}\url{#1}}%
}

\begin{document}
\title{Testing halo models for constraining astrophysical feedback with multi-probe modeling: I. 3D Power spectra and mass fractions}
\shorttitle{Constraining astrophysical feedback with multi-probe modeling}

\author{\OrcidIDName{0000-0003-3714-2574}{Pranjal R. S.}$^\star$}
\affiliation{Department of Astronomy/Steward Observatory, University of Arizona, 933 North Cherry Avenue, Tucson, AZ 85721, USA}
\email{$^{\star}$\textcolor{magenta}{pranjalrs@arizona.edu}}

\author{{Shivam Pandey}}
\affiliation{William H. Miller III Department of Physics \& Astronomy,
Johns Hopkins University, Baltimore, MD 21218, USA}

\author{\OrcidIDName{0000-0003-3312-909X}{Dhayaa Anbajagane}}
\affiliation{Department of Astronomy and Astrophysics, University of Chicago, Chicago, IL 60637, USA}
\affiliation{Kavli Institute for Cosmological Physics, University of Chicago, Chicago, IL 60637, USA}

\author{\OrcidIDName{0000-0001-8356-2014}{Elisabeth Krause}}
\affiliation{Department of Astronomy/Steward Observatory, University of Arizona, 933 North Cherry Avenue, Tucson, AZ 85721, USA}
\affiliation{Department of Physics, University of Arizona, 1118 E Fourth Street, Tucson, AZ 85721, USA}

\author{{Klaus Dolag}}
\affiliation{Max Planck Institute for Astrophysics, Karl-Schwarzschild-Str. 1, D-85741 Garching, Germany}

\begin{abstract}
Upcoming Stage-IV surveys will deliver measurements of distribution of matter with unprecedented precision, demanding highly accurate theoretical models for cosmological parameter inference. A major source of modeling uncertainty lies in astrophysical processes associated with galaxy formation and evolution, which remain poorly understood.  Probes such as the thermal and kinematic Sunyaev-Zel'dovich effects, X-rays, and dispersion measure from fast radio bursts offer a promising avenue for mapping the distribution and thermal properties of cosmic baryons. A unified analytical framework capable of jointly modeling these observables is essential for fully harnessing the complementary information while mitigating probe-specific systematics. In this work, we present a detailed assessment of existing analytical models, which differ in their assumptions and prescriptions for simultaneously describing the distribution of matter and baryons in the universe. Using the \textsc{Magneticum} hydrodynamical simulation, we test these models by jointly analyzing the 3D auto- and cross-power spectra of the matter and baryonic fields that underpin the above probes. We find that all models can reproduce the power spectra at sub-percent to few-percent accuracy, depending on the tracer combination and number of free parameters. Their ability to recover underlying halo properties, such as the evolution of gas abundance and thermodynamic profiles with halo mass, varies considerably. Our results suggest that these models require further refinement and testing for reliable interpretation of multi-wavelength datasets.
\end{abstract}




\section{Introduction}
The large-scale structure (LSS) of the universe forms through the gravitational collapse of matter over cosmic time. As light emitted from galaxies and the cosmic microwave background (CMB) travels through the intervening structure to our telescopes, it gets coherently distorted, offering a unique way of observing the distribution of dark and luminous matter in the universe. Tracing the growth of LSS through clustering of matter observed using weak gravitational lensing provides a powerful way to constrain the underlying cosmological model. Upcoming stage-IV observatories \citep{ Ade:2019:JCAP:, DESI2022, Mellier2025, Ivezic:2019:ApJ:, ROTAC2025} will create highly precise maps of the projected matter distribution and will require theoretical models that are accurate at the percent level to robustly perform cosmological inferences using them \citep{ Laureijs:2011:arXiv:, Hearin:2012:JCAP:, TheLSSTDarkEnergyScienceCollaboration:2018:arXiv:, EuclidCollaboration:2024:A&A:}.

High energy events such as  supernovae and feedback from supermassive black holes (active galactic nuclei, or AGN) alter the LSS even at several mega-parsecs scales. Poorly understood physics of this baryonic feedback (AGN and supernovae) is the leading source of systematics in developing percent-level precise theoretical models of matter clustering \citep{Zentner:2008:PhRvD:, Semboloni:2011:MNRAS:, vanDaalen:2011:MNRAS:,Chisari:2019:OJAp:, Huang:2019:MNRAS:}. Directly observing the distribution and thermodynamics of the baryonic component of total matter provides a potential way forward to constraining baryonic feedback.

The scattering of CMB light off the free electrons in the LSS creates observable signatures that are sensitive to the thermodynamics of the baryonic component \citep{Sunyaev:1970:CoASP:, Sunyaev:1972:CoASP:, Sunyaev:1980:ARA&A:, Sunyaev:1980:MNRAS:}. The thermal Sunyaev–Zel’dovich (tSZ) effect arises from the random thermal motion of electrons and probes the line-of-sight integrated electron pressure, while, the kinetic Sunyaev–Zel’dovich (kSZ) effect is caused by the bulk motion of electrons and is sensitive to  the velocity-weighted electron density. These maps of integrated baryonic thermodynamics can be isolated as a function of redshift by cross-correlating them with galaxy/cluster catalogs \citep[e.g.,][]{PlanckCollaboration:2013:A&A:, Greco:2015:ApJ:, Hill:2018:PhRvD:, Pandey:2020:PhRvD:,  Dalal2025,Pandey2025} and their joint modeling with other tracers provides a powerful way for constraining feedback processes \citep{Battaglia_17, Fang2024, Mead2020, Bigwood2024,  Pandey2024, To2024,Kovac2025,Schneider2025}. The hot electrons residing in massive halos also emit radiation via the Bremsstrahlung process, which can be observed using X-ray observatories \citep{Voit:2005:RvMP:, Ettori:2009:A&A:, Arnaud:2011:HEAD:, Ghirardini:2024:A&A:}. These observations probe the integrated temperature and density of the gas and provide crucial information about the impact of feedback processes \citep{Singh2016,Schneider2022,Ferreira2024,Grandis2024, Popesso2024,Posta2024}. Finally, observables such as dispersion measure of fast radio bursts (FRBs) can also provide complementary information for mapping the gas distribution \citep[e.g.][]{Nicola2022, Batten2022,  Reischke2023, Theis2024, Medlock2025,Sharma2025}. 

Hydrodynamical simulations such as:  CAMELS \citep{Navarro2021}, IllustrisTNG \citep{Nelson2019}, MilleniumTNG \citep{Pakmor2023}, FLAMINGO \citep{Schaye2023}, Magneticum \citep{Dolag2025}, are the most physically accurate way we currently have to model galaxy formation, astrophysical feedback, and its impact on the large-scale structure (e.g., see \citealt{Wechsler:2018:ARA&A:} for a review).  These simulations use effective sub-grid physics to control the properties of baryonic feedback mechanisms and simultaneously evolve all components of the LSS. Therefore, cross-correlations between fields (e.g., electron density, pressure, and total matter) are consistently captured in hydrodynamical simulations, conditioned on the assumed sub-grid physics, which is often calibrated using a subset of observations. Cosmological simulations remain computationally expensive, with single state-of-the-art runs covering only $\sim 0.5 {\rm Gpc}^3$ and taking millions of CPU-hours. This makes it impractical to explore the model space of sub-grid physics parameters using Bayesian methods with hydrodynamical simulations. This provides a strong motivation to develop analytical models that capture the impact of astrophysical feedback on baryonic thermodynamics and distribution of matter components.

The halo model \citep[see][for a review]{cooray_halo_2002,Asgari2023} provides a natural framework to develop analytical models that incorporate the impact of baryonic feedback on the LSS \citep[e.g.,][]{Mead2020,Mead2021,Pandey2024}. 
In this work, we benchmark analytical models by jointly fitting simulation measurements of the auto-power spectra of total matter distribution, cross-power spectrum between matter density and electron pressure, and the power spectrum of the electron density field. The first two power spectra directly source the observable weak lensing  and shear-tSZ signals, respectively. The electron density power spectrum is key in interpreting observables sensitive to the distribution of free electrons, such FRB dispersion measures and the kSZ effect. We use the parameter constraints to infer the abundance of bound gas and stars as well as thermodynamic profiles of electron pressure, temperature, and density, as a function of halo mass and test their accuracy against simulation measurements.
 This work is among the first steps in assessing whether analytical models are sufficiently flexible to jointly and self-consistently model these observables. Such an assessment is essential not only for reliable cosmological inference from multi-probe analysis but also for developing a more comprehensive understanding of astrophysical processes in the late universe.

We focus on three models from the literature that incorporate correlated evolution of matter and baryonic components, albeit making different assumptions regarding the behavior of components of interest: Mead20+ \citep{Mead2020}, Arico24+ \citep{Arico2024}, Schneider19+ \citep{Schneider2019, Pandey2024, Anbajagane2024}\footnote{In this work we use updated variants of the Mead, Arico, and Schneider models, which we denote with a “+”. We refer the reader to Sec. \refc{sec:halo_models} for details.}. We use the Magneticum simulation \citep{Dolag2025}, which was not used in the calibration of either model considered in this paper, as an independent test of model flexibility in reproducing various cross-correlations. We also assess model performance by inferring physical properties of baryons in halos, such as gas and stellar mass fractions.

The paper is organized as follows: In Sec.~\refc{sec:halo_models} we introduce the halo models and detail their modeling ingredients and assumptions. In Sec.~\refc{sec:model_comparison} we assess the correlations between model parameters, how degeneracies are broken when including different power spectra, as well as a qualitative comparison of flexibility across different models. In Sec.~\refc{sec:simulation} we describe the properties of the \textsc{Magneticum} simulations used in this work. In Sec.~\refc{sec:analysis_setup} we describe the analysis methodology. In Sec.~\refc{sec:results} we present the results from fitting models to measurements from the hydrodynamical simulation. We analyze their accuracy in fitting various combinations of power spectra and also examine the inferred constraints on halo properties and scaling relations. We present our conclusions and outlook for future work in Sec.~\refc{sec:discussion}.

\section{Halo Model Implementations}
\label{sec:halo_models}

The halo model is a formalism for predicting the statistics of cosmological fields by positing that all matter resides in spherically collapsed over-densities called halos \citep[e.g.][for a recent review]{Asgari2023}. The clustering on large scales is dominated by the correlation across halo pairs which is captured by the two-halo term,  while on small scales it is determined by the distribution of the field within individual halos aka. the one-halo term.

For any two fields u$(\boldsymbol{r})$ and v$(\boldsymbol{r})$, the two- and one- halo terms are expressed as 

\begin{align}
\label{eq:2h_term}
    P_{\rm uv}^{\rm 2h}(k) &= P_{\rm lin}(k)\prod_{i=\rm u,v}\left[ \int_0^\infty b(M) n(M) \hat{W}_i(k,M) {\rm d}M \right],\\
    P_{\rm uv}^{\rm 1h}(k) &= \int_0^\infty n(M) \hat{W}_{\rm u}(k,M) \hat{W}_{\rm v}(k,M) {\rm d}M. \label{eq:1h_term}
\end{align}

Here $P_{\rm lin}(k)$ is the linear matter power spectrum of matter fluctuations, $M$ is the halo mass, $b(M)$ is the linear halo bias,  and $n(M)$ is the comoving number density of halos per unit mass. In the above expression, $W_{\rm u}(r,M)$ denotes the halo profile of the field whose power spectrum is being calculated. For instance, when computing the matter power spectrum, $W_{\rm u}(r,M)=\rho_{\rm m}(r,M)/\bar{\rho}_{\rm m}$, where $\rho_{\rm m}(r,M)$ is the total matter density profile of a halo of mass $M$ and $\bar{\rho}_{\rm m}$ is the mean matter density of the universe. With the assumption of spherical symmetry, the Fourier equivalent of the real space profile $W_{\rm u}(r,M)$ is obtained as
\begin{equation}
    \hat{W}_{\rm u}(k, M) =   \int_0^\infty \frac{\sin (kr)}{kr} W_{\rm u}(r, M) 4\pi r^2{\rm d}r.
\end{equation}

The integrals in Eqs.~(\refc{eq:2h_term}) and (\refc{eq:1h_term}) are typically evaluated over a finite halo mass range, from $M_1$ to $M_2$. Because the halo mass function drops steeply at the high-mass end, choosing a sufficiently large $M_2$ generally ensures good convergence. In contrast, there is a significant amount of mass contained in low mass halos and therefore the integral converges much more slowly as $M_1$ decreases. To account for the contribution from halos below $M_1$, the two-halo term receives an additive correction  which ensures that $P_{\rm uv}^{\rm 2h}(k\rightarrow0)=P_{\rm lin}(k\rightarrow0)$ \citep{Cacciato2012,Schmidt2016}. This correction assumes that the cumulative contribution of halos below $M_1$ can be approximated as originating from halos with mass exactly equal to $M_1$. In our calculations, we set  $M_1=10^{10}\,{\rm M}_\odot$ and $M_2=10^{16}\,{\rm M}_\odot$.

A common challenge for the halo model is the
inaccuracy on scales where the power spectrum transitions from being dominated by the two-halo term to the one-halo term. It has been shown the lack of power is primarily a result of adopting a linear halo bias and almost absent when a full non-linear halo bias prescription is used \citep{Mead2021b} and is related to the halo exclusion effects \citep{Zehavi:2004:ApJ:, Tinker2005}. To model this transition regime, we combine the two terms as \citep{Mead2015,Mead2021}
\begin{equation}
\label{eq:Pk_smooth}
    P_{\rm uv}(k) = \left[(P_{\rm uv}^{\rm 1h}(k))^{\alpha_\text{smooth}} + (P_{\rm uv}^{\rm 2h}(k))^{\alpha_\text{smooth}}\right]^{1/\alpha_\text{smooth}},
\end{equation}
where $\alpha_\text{smooth}$ is a free parameter. We further discuss this in Sec.~\refc{sec:analysis_setup}.

\noindent\textbf{Power spectrum response}
As we later discuss in Sec.~\refc{sec:analysis_setup}, we compare model predictions of the \textit{power spectrum response} to measurements from hydrodynamical simulations. The power spectrum response is defined as 
\begin{equation}
\label{eq:response}
    R_{\rm uv}(k) = \frac{P_{\rm uv}(k)}{P_{\rm mm,dmo}(k)},
\end{equation}
where $P_{\rm mm,dmo}(k)$ is the matter power spectrum when all the matter in a universe is in the form of dark matter. For the Mead20+ (Sec.~\refc{sec:M20}) and Arico24+ (Sec.~\refc{sec:A24}) model, $P_{\rm mm,dmo}(k)$ is computed assuming that the halo matter density is well described by an NFW profile
\begin{equation}
\label{eq:nfw}
    \rho_{\rm nfw}(r) = \frac{\rho_{\text{nfw},0}}{\left(\frac{r}{r_\text{s}}\right)\left(1+\frac{r}{r_\text{s}}\right)^2},
\end{equation}
where $\rho_{\text{nfw},0}$ is a normalization constant and $r_\text{s}=r_\text{200c}/c_\text{200c}$ is the scale radius. On the other hand, we use a truncated NFW profile for the Schneider19+ model (Sec.~\refc{sec:S19})
\begin{equation}
\label{eq:tnfw}
    \rho_\text{tnfw}(r) = \frac{\rho_{\text{tnfw},0}}{\left(\frac{r}{r_\text{s}}\right)\left(1+\frac{r}{r_\text{s}}\right)^2}\frac{1}{\left(1+\left(\frac{r}{r_\text{t}}\right)^2\right)^2},
\end{equation}
where  and $r_\text{t}=4\times r_\text{200c}$ is the truncation radius \citep{Oguri2011,Schneider2019}.

All modeling in this work is done using the \textsc{BaryonForge}\footnote{\murl{https://github.com/DhayaaAnbajagane/BaryonForge}} pipeline \citep{Anbajagane2024}, and associated halo model routines provided in the \textsc{Core Cosmology Library} (CCL)\footnote{\murl{https://github.com/LSSTDESC/CCL}} \citep{Chisari2019}. We have verified that the predictions of the individual profiles, described in the sections below, from \textsc{BaryonForge} reproduce the predictions from original studies.

We adopt the halo bias and mass function from \citet{Tinker2010} and use the mass–concentration relation from \citet{Diemer2015}. Additionally, we define halos as collapsed structures with an average density 200 times the critical density of the universe. The halo radius, 
$r_{\rm 200c}$, and its mass, 
$M_{\rm 200c}$, are related by $M_{\rm 200c}=4/3\pi r_{\rm 200c}^3 \rho_{\rm crit}$, where $\rho_{\rm crit}$ is the critical density. The linear matter power spectrum is computed using the \textsc{Camb} Boltzmann code \citep{Lewis2000,Lewis2011}. We assume a $\Lambda$CDM cosmology consistent with the fiducial cosmology parameters of the \textsc{Magneticum} simulation (Sec.~\refc{sec:simulation}): $\Omega_\text{m}=0.272, \Omega_\text{b}=0.0456,  \Omega_\Lambda=0.728, \sigma_8=0.809, h=0.704$, and  $n_\text{s}=0.963$.

In the following sections, we summarize the main ingredients for the models considered in this study: Mead20+, Schneider19+, and Arico24+. The model parameters are summarized in Table~\refc{tab:model_params} and a detailed description of the three models is provided in Appendix~\refc{appx:models}.

\subsection{Mead20+}
\label{sec:M20}
\citet{Mead2020} (hereafter \citetalias{Mead2020}) were among the first to develop a halo model for jointly analyzing weak lensing and tSZ data, with an emphasis on a physical link between the fields sourcing the two signals. The model was calibrated using the BAHAMAS simulations and has been used to jointly analyze weak lensing and tSZ data \citep{Tilman2022}.

The primary assumption that underpins the model is that the impact of baryonic feedback on halo structure can be captured by modifying the halo concentration (Eq.~\refc{eq:HMx_cmod}). It adopts simple prescriptions for describing the spatial distribution of gas and non-thermal pressure support, therefore it has fewer free parameters compared to the other models considered in this work. Specifically, the model uses four parameters to describe the bound and ejected gas profiles. The bound gas fraction is set by a two-parameter scaling relation, while the stellar abundance is governed by four parameters. The ejected gas component is computed as the difference between the cosmic baryon fraction and the sum of bound gas and stellar mass fraction. Gas pressure is calculated using the ideal gas law combined with a polytropic equation of state. Non-thermal pressure support is modeled as a global amplitude rescaling of the pressure profile, independent of halo mass and radius. A full list of model parameters is provided in Table~\refc{tab:model_params}.

In this work we use an updated version of the model, referred to as Mead20+, which, among other improvements, incorporates a more physically motivated prescription for the ejected gas component. Further details are provided in Appendix~\refc{appx:M20}.

\begin{table*}
    \centering
    \begin{tabular}{ccllll}
    \toprule
    Parameter & Equation & \multicolumn{1}{c}{Description} & & \multicolumn{1}{c}{Prior}& \multicolumn{1}{c}{Fiducial}\\
    \hline
    \multicolumn{5}{c}{\textbf{Mead20+}} \\
    \hspace{5.5pt}$\epsilon_1$ \textcolor{Mahogany}{$^\dagger$} & \refc{eq:HMx_cmod} & Modification in halo concentration for gas-poor halos &\hspace{-1em}\rdelim\}{6}{*}[{} Gas]& $\mathcal{U}(-0.6, 0.6)$& -0.11\\
    $\epsilon_2$ &~\refc{eq:HMx_cmod} &Modification in halo concentration for gas-rich halos  & & $\mathcal{U}(-0.6, 0.6)$& 0.0\\
    \hspace{5.5pt}$\Gamma$ \textcolor{Mahogany}{$^\dagger$} &~\refc{eq:HMx_BG}  &Polytropic index for the equation of state of bound gas in halos & & $\mathcal{U}(1.05, 3)$& 1.17\\
    \hspace{5.5pt}$\log (M_0h/\text{M}_\odot)$  \textcolor{Mahogany}{$^\dagger$} &\refc{eq:HMx_fBG}  &Halo mass for which halos have lost half of their initial gas content & & $\mathcal{U}(10, 16)$ & 13.59\\
    $\beta$ &~\refc{eq:HMx_fBG} &Slope of the bound gas fraction-halo mass relation & & $\mathcal{U}(0.1, 2)$ & 0.6 \\
    $\eta_\text{b}$ &~\refc{eq:HMx_etab} & Maximum distance to which gas is ejected & & $\mathcal{U}(0.2, 5)$ & 0.5\\ 
    \vspace{\baselineskip}\\
    $A_*$ &~\refc{eq:HMx_fstar} &Peak fraction of halo mass that is in stars &\hspace{-1em}\rdelim\}{4}{*}[{} Stars] &$\mathcal{U}(0.02, 0.08)$ & 0.03\\
    $\log (M_*h/\text{M}_\odot)$ &\refc{eq:HMx_fstar} &Halo mass at which star-formation efficiency peaks& & $\mathcal{U}(10, 14.5)$  & 12.45\\
    $\sigma_*$ &~\refc{eq:HMx_fstar} &Logarithmic width of star-formation efficiency distribution&  & Fixed & 1.20\\
    $\eta$ &~\refc{eq:HMx_fcga} &Power-law index for central–satellite galaxy split & & Fixed & -0.36\\
    \vspace{\baselineskip}\\
   \hspace{5.5pt}$\alpha$ \textcolor{Mahogany}{$^\dagger$} &~\refc{eq:HMx_Tv}  &Ratio of halo temperature to that of virial equilibrium & \hspace{-1em}\rdelim\}{2}{*}[{} Pressure] & $\mathcal{U}(0, 1)$ & 0.85\\
   \hspace{5.5pt}$\log (T_\text{w}/$K) \textcolor{Mahogany}{$^\dagger$}& \refc{eq:HMx_Pe} &Temperature of ejected gas & & $\mathcal{U}(6, 7.5)$& 6.65\\

    \hline \multicolumn{5}{c}{\textbf{Schneider19+}}\\
    $\gamma$ &~\refc{eq:god_rhogas}& Outer slope of bound gas profile & \hspace{-1em}\rdelim\}{11}{*}[{} Gas]& $\mathcal{U}(1, 6)$ & 2\\
    \hspace{3.5pt}$\delta$ \textcolor{Mahogany}{$^\dagger$}&~\refc{eq:god_rhogas}& Outer slope of bound gas profile & & $\mathcal{U}(2, 11)$ & 7\\
    $\log (M_\text{c,0}h/\text{M}_\odot)$  &~\refc{eq:god_beta}&Halo mass below which gas profile becomes shallower than NFW & & $\mathcal{U}(11, 16)$& 14.83\\
    \hspace{3.5pt}$\mu_\beta$\textcolor{Mahogany}{$^\dagger$} &~\refc{eq:god_beta}& Mass dependence of inner slope $\beta$ & &$\mathcal{U}(0.0, 2.0)$ & 0.6\\
        \vspace{\baselineskip}\\

    $\theta_\text{co,0}$ &~\refc{eq:god_theta}& Radius at which the inner slope of the gas density profile changes & & $\mathcal{U}(10^{-3}, 0.5)$ & 0.05\\
        $\log (M_\text{co}h/\text{M}_\odot)$  &~\refc{eq:god_theta}& Pivot for evolution of $\theta_\text{co,0}$ with halo mass & & Fixed &  16\\
    $\mu_\text{co}$ &~\refc{eq:god_theta}& Power law index for mass dependence of radius parameter & & $\mathcal{U}(-1.0, 1.0)$& 0.0\\

    \hspace{3.5pt}$\theta_\text{ej,0}$\textcolor{Mahogany}{$^\dagger$} &~\refc{eq:god_theta}& Radius at which the outer slope of the gas density profile changes & & $\mathcal{U}(1.0,8.0)$ & 2\\
$\log (M_\text{ej}h/\text{M}_\odot)$  &~\refc{eq:god_theta}&  Pivot for evolution of $\theta_\text{ej,0}$ with halo mass & & Fixed & 16\\
    $\mu_\text{ej}$ &~\refc{eq:god_theta}& Power law index for mass dependence of radius parameter & &$\mathcal{U}(-2.0, 2.0)$ &   0.0\\
        \vspace{\baselineskip}\\

    $A$  &~\refc{eq:god_fstar} & Peak fraction of halo mass that is in stars & \hspace{-1em}\rdelim\}{6}{*}[{} Stars] & $\mathcal{U}(0.01, 0.07)$& 0.045\\
    $\log (M_1h/\text{M}_\odot)$  &~\refc{eq:god_fstar} & Characteristic halo mass for central galaxy & & Fixed & 11.5\\
    \hspace{3.5pt}$\eta$\textcolor{Mahogany}{$^\dagger$}  &\refc{eq:god_fstar} & Power-law scaling of total stellar fraction for $M_\text{200c}\gg M_1$ & &$\mathcal{U}(0.05, 0.5)$ & 0.3\\
    $\eta_\delta$  &\refc{eq:god_fcga} & Power-law index for central galaxy at high-mass end $\eta_{\rm cga}=\eta+\eta_{\rm \delta}$ & & $\mathcal{U}(0.05, 0.5)$& 0.1\\
    $\tau$  &\refc{eq:god_fstar} & Power-law scaling of total stellar fraction for $M_\text{200c}\ll M_1$ & & Fixed & -1.5\\
    $\tau_\delta$  &\refc{eq:god_fcga} & Power-law index for central galaxy at low-mass end $\tau_{\rm cga}=\tau+\tau_{\rm \delta}$ & &  Fixed & 0.0 \\
   
        \vspace{\baselineskip}\\

    \hspace{3.5pt}$\alpha_\text{nt}$\textcolor{Mahogany}{$^\dagger$} &~\refc{eq:god_Rnt} & Non-thermal pressure fraction at $R_{\rm 200c}$ & \hspace{-1em}\rdelim\}{2}{*}[{} Pressure] & $\mathcal{U}(0.0, 0.5)$&0.18 \\
    $\gamma_\text{nt}$ &~\refc{eq:god_Rnt} & Power-law scaling of non-thermal pressure with radius & & $\mathcal{U}(0.0, 4.0)$ & 0.3\\

    \hline\multicolumn{5}{c}{\textbf{Arico24+}}\\

    \hspace{3.5pt}$\theta_\text{inn}$\textcolor{Mahogany}{$^\dagger$} &~\refc{eq:AA_rhoBG} &Radius at which the inner slope of the gas  profile changes& \hspace{-1em}\rdelim\}{9}{*}[{} Gas] & $\mathcal{U}(10^{-3}, 4.0)$& 0.3\\
     $\theta_\text{out}$ &~\refc{eq:AA_rhoBG} &Radius at which the outer slope of the gas  profile changes & & $\mathcal{U}(0.8, 10.0)$& 1.0\\
     $\log (M_\text{inn}h/\text{M}_\odot)$ &~\refc{eq:AA_betai} &Halo mass for which inner slope matches the NFW profile & & $\mathcal{U}(10, 16)$& 12\\
     $\mu_\text{i}$ &~\refc{eq:AA_betai}& Mass dependence of inner  slope of gas profile & & $\mathcal{U}(0, 1)$& 0.15\\
     \hspace{3.5pt}$\log (M_\text{c}h/\text{M}_\odot)$\textcolor{Mahogany}{$^\dagger$} &~\refc{eq:AA_fHG} & Halo mass for which halos have lost half of their initial gas content &  & $\mathcal{U}(10, 16)$& 13.0\\
     $\beta$ &~\refc{eq:AA_fHG} & Slope of the bound gas fraction-halo mass relation & & $\mathcal{U}(0.1, 5)$& 0.35\\
      
     $\eta$& Above~\refc{eq:AA_rho_RG} &Physical extent of ejected gas & & $\mathcal{U}(0.2, 5)$& 0.5\\ 
     
     $\log (M_\text{r}h/\text{M}_\odot)$ &~\refc{eq:AA_fRG} & Mass pivot for determining the reacrreted gas fraction & & Fixed & 18.0\\
     $\beta_\text{r}$&~\refc{eq:AA_fRG} & Slope of the reacrreted gas fraction to halo mass relation & & Fixed& 2 \\
         \vspace{\baselineskip}\\

  \hspace{3.5pt}$\log (M_{1,0}h/\text{M}_\odot)$\textcolor{Mahogany}{$^\dagger$} &  Below~\refc{eq:AA_fcga} &Characteristic halo mass scale for central galaxies & \hspace{-1em}\rdelim\}{6}{*}[{} Stars] &$\mathcal{U}(10, 14.5)$ & 12.0\\
    $M_{1,\text{sat}}$ & Below~\refc{eq:AA_rhocga} & Scaling between satellite and central galaxy parameters & & $\mathcal{U}(0, 5)$& 3.98\\
    $\epsilon_\text{sat}$ & Below~\refc{eq:AA_rhocga} & Scaling between satellite and central galaxy parameters & & $\mathcal{U}(0, 2.5)$& 1.0\\
    $\alpha_\text{sat}$ & Below~\refc{eq:AA_rhocga} & Scaling between satellite and central galaxy parameters & & Fixed & 1.0\\
    $\delta_\text{sat}$ & Below~\refc{eq:AA_rhocga} & Scaling between satellite and central galaxy parameters & & $\mathcal{U}(0, 2.5)$& 0.99\\
    $\gamma_\text{sat}$ & Below~\refc{eq:AA_rhocga} & Scaling between satellite and central galaxy parameters & & $\mathcal{U}(0, 2.5)$& 1.67\\
      \vspace{\baselineskip}\\

    \hspace{3.5pt}$A_{\rm th}$\textcolor{Mahogany}{$^\dagger$} &~\refc{eq:AA_Ant} & Fraction of thermal pressure support & \hspace{-1em}\rdelim\}{2}{*}[{} Pressure] & $\mathcal{U}(0, 5)$ & 0.495\\
    \hspace{3.5pt}$\log(T_{\rm w}/{\rm K})$\textcolor{Mahogany}{$^\dagger$} &Below ~\refc{eq:AA_Ant} & Temperature of ejected gas &  & $\mathcal{U}(6, 7.5)$ & 6.5\\
    \bottomrule
    \end{tabular}
    \caption{Summary of halo model parameters. The table also shows the priors for all parameters varied when fitting simulation measurements, we refer to this as the \textbf{extended} set. Only the five parameters marked by \textcolor{Mahogany}{$^\dagger$} are varied in the \textbf{six-parameter} set while the rest are held fixed at their fiducial values. Both parameter sets for each model also include an additional parameter $\alpha_{\rm smooth}$ to model the transition regime.}
    \label{tab:model_params}
\end{table*}

\subsection{Schneider19+}
\label{sec:S19}
The \textit{baryonification} method introduced in \citet{Schneider2016,Schneider2019} uses analytical recipes for modeling the impact of astrophysical feedback on the density of matter components (dark and baryonic) in halos and has been widely used to constrain feedback effects from observational data \citep{Giri2021,Schneider2022,Grandis2024, Bigwood2024}. This method has been extended, which we refer as Schneider19+, for joint analysis with SZ surveys by also modeling the thermodynamic properties of the gas distribution \citep{Anbajagane2024,Pandey2024}. 

The gas is modeled as a single component with a highly flexible density profile, specified by 10 parameters. Its gravitational back-reaction on the dark matter halo is included by assuming adiabatic relaxation (Eq.~\refc{eq:god_relax}). Both the gas and dark matter profiles are untruncated and extend beyond the nominal halo boundary. The total gas abundance is set by subtracting the stellar mass fraction, parameterized by a six-parameter scaling relation, from the cosmic baryon fraction. Gas pressure is obtained by solving the equation of hydrostatic equilibrium, instead of relying on a fixed equation of state. Non-thermal pressure is included using the \citet{Shaw2010} model, which introduces two additional free parameters. The parameters are summarized in Table~\refc{tab:model_params} and the modeling ingredients are detailed in Appendix~\refc{appx:S19}.

\subsection{Arico24+}
\label{sec:A24}
The methodology presented in \citet{Arico2024} (hereafter, A24) builds on earlier baryon correction models \citep{Schneider2015, Schneider2019, Arico2020, Arico2021}, extending them to include thermodynamic quantities such as temperature and electron pressure for applications to tSZ observations. While originally developed as a baryonification approach, i.e., designed to modify particle positions in \textit{N}-body simulations, the use of analytical prescriptions makes it readily adaptable in a halo model framework.

The model assumes that the gravitational back-reaction from redistributed baryons can be captured by modeling the dark matter halo's response as adiabatic relaxation. The gas distribution is described using a flexible five-parameter profile and is divided into three components: bound, reaccreted, and ejected gas. The abundances of the bound and reaccreted components are set by separate scaling relations, each with two free parameters. The stellar mass fraction is specified using a six-parameter relation. Gas pressure is calculated assuming a polytropic equation of state, and the non-thermal pressure contribution is included via the calibrated prescription from \citet{Green2020}. 

We also include an additional parameter $T_{\rm w}$ which sets the temperature of the ejected gas. In our initial tests we found that this parameter was important for obtaining good fits to the matter--pressure power spectrum --- we refer to this updated model as Arico24+. A detailed model description can be found in Appendix~\refc{appx:A20} and the model parameters are listed in Table~\refc{tab:model_params}.

\section{Model comparison}
\label{sec:model_comparison}
In this section we present a qualitative comparison of the different models. To build intuition for the respective parameter space of the different halo model implementations, we present here a qualitative comparison of the different model spaces. First, we examine the correlations between parameters and how these change depending on which power spectrum is being analyzed. This highlights how different observables constrain different physical components of the models. We then compare their prediction space, the collection of all possible model outputs, to understand their relative flexibility.

\subsection{Parameter degeneracies}
\begin{figure*}
    \centering
    \includegraphics[width=\linewidth]{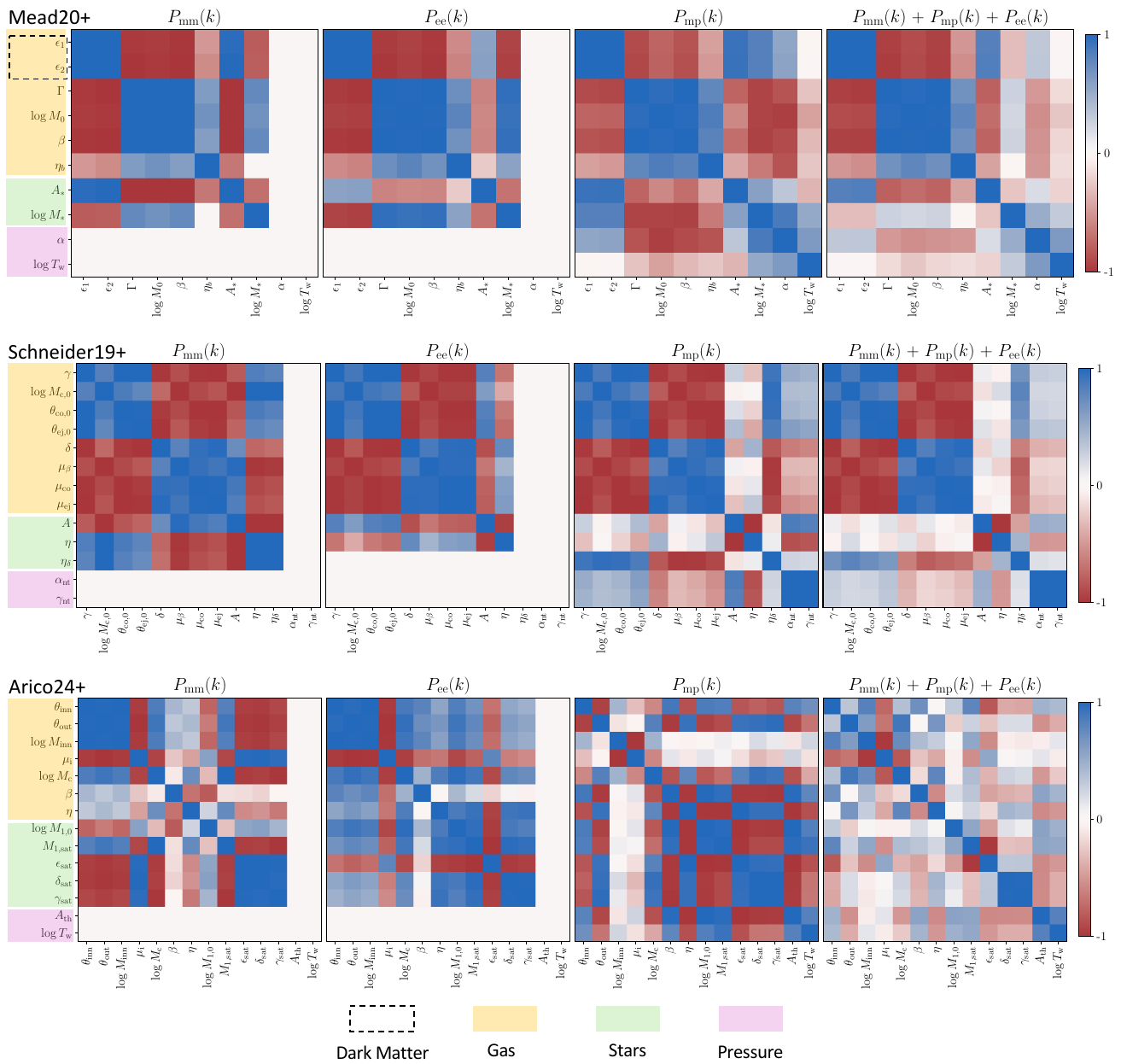}
    \caption{Correlations between halo model parameters for combinations of different matter fields. \textbf{Left to right:} Parameter correlations for matter--matter, electron density--electron density, matter--pressure  3D power spectra and when analyzing all of them combined. \textbf{Top to bottom:} Parameter correlations for the Mead20+, Schneider19+, and Arico24+ models, respectively. We see that combining information from different probes helps in weakening parameter correlations, however, the degree of decorrelation varies across models. Note that the parameters $\{\alpha,T_\text{w} \}$ do not impact $P_{\rm mm}(k)$ and $P_{\rm ee}(k)$, hence they are not correlated with the other model parameter for the Mead20+ model. The same applies to  $\{\alpha_\text{nt}, \gamma_\text{nt}\}$ in the case of Schneider19+ and $\{A_{\rm th},T_\text{w}\}$ for Arico24+.}
    \label{fig:corr_matrix}
\end{figure*}

\begin{figure*}
    \centering
    \includegraphics[width=0.326\linewidth]{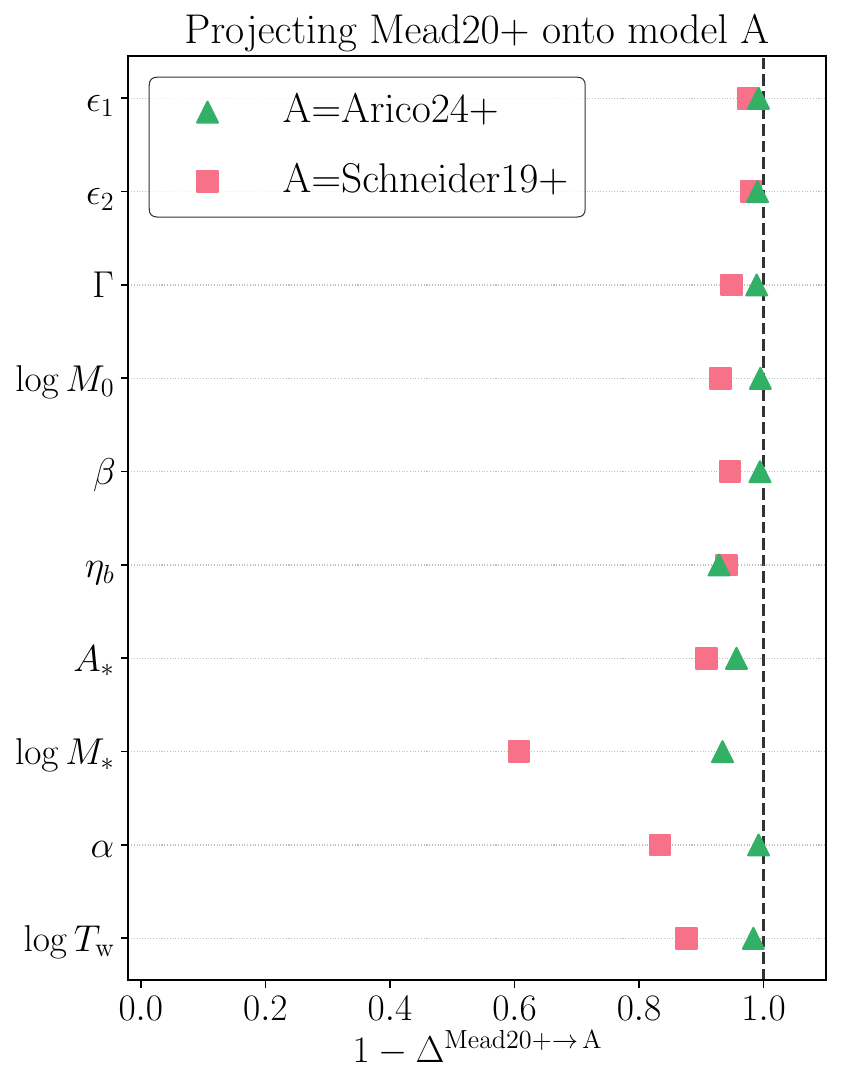}
    \includegraphics[width=0.333\linewidth]{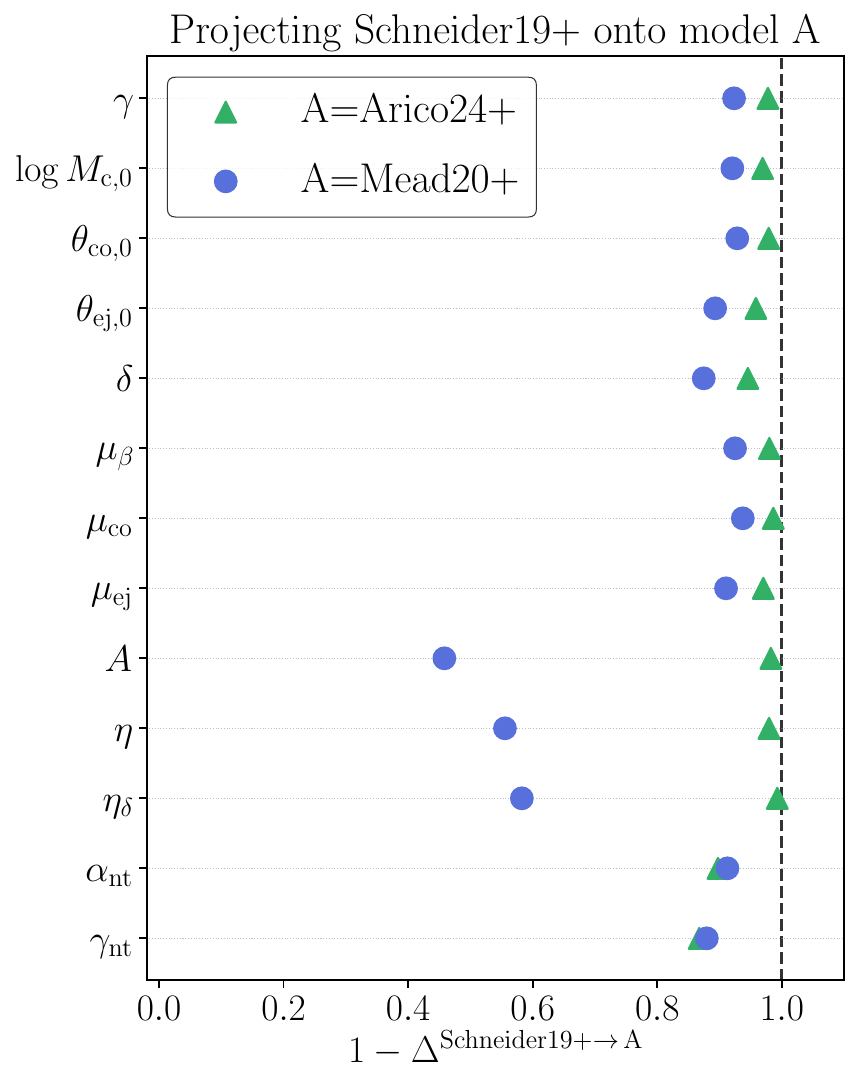}
    \includegraphics[width=0.334\linewidth]{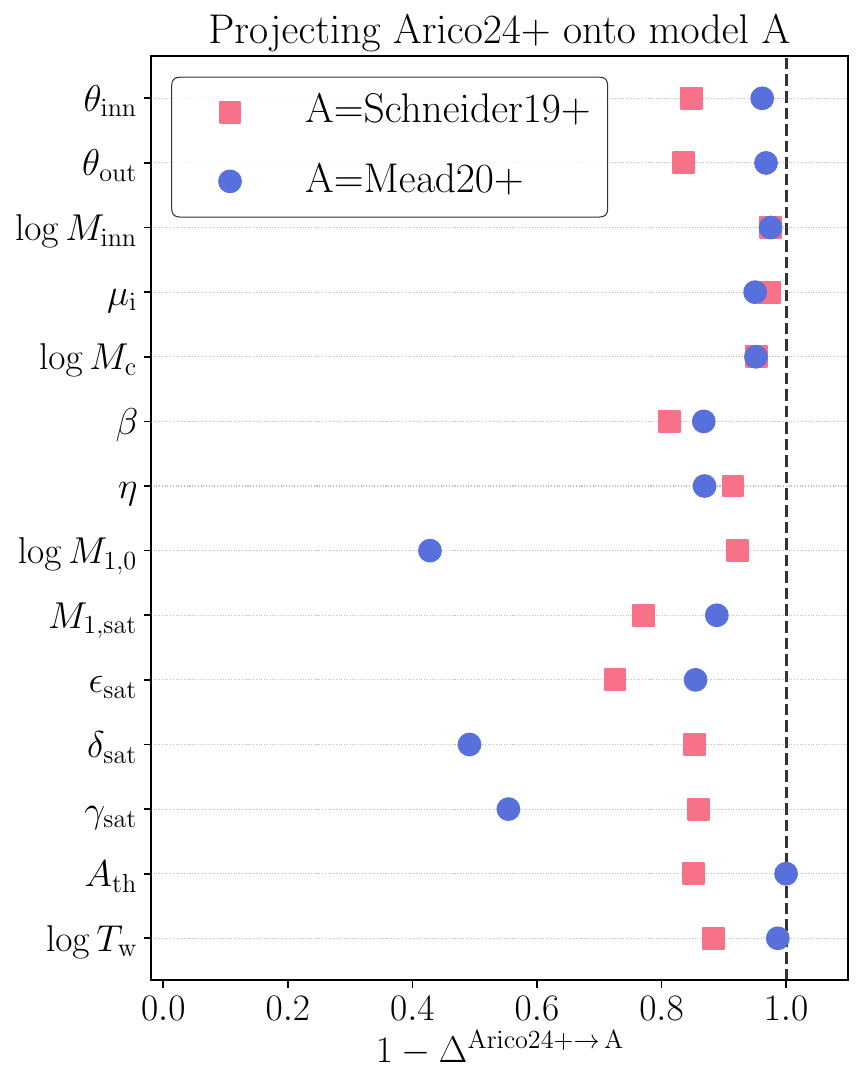}

    \caption{Projection of parameters across halo model implementations. Each panel illustrates how well the impact of a single parameter on the concatenated data vector ($P_{\rm mm}$, $P_{\rm ee}$, $P_{\rm mp}$) can be represented within other models. The dashed black line denotes when a parameter can be perfectly represented. Panels from left two right show projections of the Mead20+, Schneider19+, and Arico24+ models, respectively.}
    \label{fig:basis_proj}
\end{figure*}

It is instructive to examine how model parameters correlate with one another and particularly how these correlations evolve when different observables are included in the analysis. Exploring the structure of the parameter space and identifying major degeneracies can guide future efforts to refine these models.

For a model A, we compute the correlations between parameters $\vartheta_i^\text{A}$ and $\vartheta_j^\text{A}$ as the dot product of their respective derivative vectors
\begin{equation}
    \text{Corr}(\vartheta_i^\text{A}, \vartheta_j^\text{A}) = \langle \boldsymbol{\Theta}^\text{A}_i \cdot \boldsymbol{\Theta}^\text{A}_j\rangle,
\end{equation}
where,
\begin{equation}
\label{eq:dPk_dtheta}
    \boldsymbol{\Theta}^\text{A}_i=\frac{\partial \log P_{\rm uv}(k)}{\partial \log \vartheta_i^\text{A}}\, \, \text{and}\, \,  ||\boldsymbol{\Theta}^\text{A}_i||=1.
\end{equation}
Each vector $\boldsymbol{\Theta}^\text{A}_i$ represents how the power spectrum responds to variations in the parameter $\vartheta_i^\text{A}$, effectively defining the directions in prediction space that are captured by the model. The power spectrum $P_{\rm uv}(k)$ can correspond to either auto-correlations of a single field (e.g., matter–matter $P_{\rm mm}$ or electron density--electron density $P_{\rm ee}$) or cross-correlations between different fields (e.g., matter–pressure $P_{\rm mp}$). One can also compute the correlation by jointly considering the change in multiple power spectra, in which case the derivative in Eq.~(\refc{eq:dPk_dtheta}) is computed for the concatenated power spectra i.e., $\{P_{\rm mm}, P_{\rm ee}, P_{\rm mp}\}$.

Figure~\refc{fig:corr_matrix} shows the correlations between model parameters when considering different power spectra. The rows, from top to bottom, show correlations for the Mead20+, Schneider19+, and Arico24+ models, respectively. The power spectra are computed for $k=0.04-10\,{\rm Mpc}^{-1}h$.

Across all models, we see that most parameters exhibit strong degeneracies when only considering an individual power spectrum i.e. either $P_{\rm mm}(k)$, $P_{\rm ee}(k)$ or $P_{\rm mp}(k)$. The parameters describing different halo components are highly correlated, and there are also degeneracies between different parameter sets, such as those describing bound gas and stellar profiles -- a feature that is common among all models. In the Mead20+ model for instance, both $\epsilon_1$ and $\epsilon_2$ change the halo concentration, albeit at different halo masses, and consequently shift the matter power spectrum in the same direction by enhancing or suppressing power in the one-halo regime. In contrast, increasing $M_0$ reduces the bound gas fraction in progressively more massive halos, which now leads to a suppression of power in the one-halo regime, an effect opposite to that of $\{\epsilon_1, \epsilon_2\}$.

Looking at the last panel in each row we see that combining different power spectra significantly reduces parameter degeneracies. In particular, the parameters governing the stellar component become largely decoupled from those describing the gas component. Since the stellar and gas contributions must together sum to the cosmic baryon fraction, increasing the stellar mass fraction reduces the available gas mass, which strongly impacts the amplitudes of $P_{\rm ee}(k)$ and $P_{\rm mp}(k)$. Similarly, the parameter governing the pressure profiles have no impact on the matter--matter or electron density power spectrum, and are naturally better constrained when the power spectra are combined. The block structure for the different models also shows that the degree of decorrelation is quite model dependent.

Overall, these results shows the importance of combining different tracers for breaking degeneracies and pinning down model parameters.

\subsection{Model feature space}
We now turn our attention towards understanding how  the prediction space of one model relates to that of another.
Since the models differ in both their assumptions and parameterizations, comparing their prediction spaces provides valuable insight into their relative and whether one model's parameter space is flexible enough to encompass the features of another.

We first define basis vectors of a given model as a set of linearly independent vectors that span its prediction space. To compare the degrees of freedom across models, we can express the basis vectors of model B in terms of model A and evaluate how accurately they are be represented. Specifically, we project the basis vectors of one model onto the orthonormal basis of another and quantify the mismatch between the original vectors and their projections.

We define the matrix $\Xi^\text{A}$ whose columns correspond to the derivative vectors of model A
\begin{equation}
    \Xi^\text{A} = 
    \begin{bmatrix}
        | & |  &   & | \\
            \frac{\partial \log P(k)}{\partial \log \vartheta_1^\text{A}} & \frac{\partial \log P(k)}{\partial \log \vartheta_2^\text{A}} & \dots & \frac{\partial \log P(k)}{\partial \log \vartheta_n^\text{A}}\\

        \vert & \vert &  & \vert
    \end{bmatrix},
\end{equation}
where $\{\vartheta_i\}$ are parameters of model A. These derivative vectors describe how the predicted power spectrum responds to parameter variations, and therefore form a natural basis for the model's prediction space. To find an orthogonal basis for model A, we perform a singular value decomposition (SVD) of $\Xi^\text{A}$
\begin{equation}
    \Xi^\text{A} = \mathbb{U}\,\, \Sigma\,\, \mathbb{V}^\text{T},
\end{equation}
where $\mathbb{U}$ and $\mathbb{V}$ are square unitary matrices and $\Sigma$ is a diagonal matrix. The first $n$ column vectors $\{\boldsymbol{\Phi}^\text{A}_1, \boldsymbol{\Phi}^\text{A}_2\dots\boldsymbol{\Phi}^\text{A}_n\}$ of $\mathbb{U}$ form an orthogonal set which spans the entire model space.

We can express a basis vector of model B, $\boldsymbol{\Theta}^\text{B}_i$, in the orthonormal basis of model A as
\begin{equation}
   \boldsymbol{\Theta}^{\text{B}\rightarrow\text{A}}_i = \sum_{j=1}^n \langle\boldsymbol{\Theta}^\text{B}_i \cdot \boldsymbol{\Phi}^\text{A}_j\rangle\,\, \boldsymbol{\Phi}^\text{A}_j
\end{equation}

We assess the accuracy of projecting model B onto model A by defining

\begin{equation}
\boldsymbol{\Delta}_i^{\text{B}\rightarrow \text{A}} = ||\boldsymbol{\Theta}^{\text{B}\rightarrow\text{A}}_i-\boldsymbol{\Theta}^\text{B}_i||,
\end{equation}

which quantifies the difference between the original data vector in model B and its projection in model A. As the value of  $\boldsymbol{\Delta}_i^{\text{B}\rightarrow \text{A}} $ approaches zero, it indicates that the variations induced by parameter $\vartheta^\text{B}_i$ in model B can be well represented by model A.

Figure~\refc{fig:basis_proj} shows the level up to which the impact of parameters on $\{P_{\rm mm}(k),\,P_{\rm ee}(k),\,P_{\rm mp}(k) \}$ can be represented in a different model. The horizontal axis shows the metric $1-\boldsymbol{\Delta}_i^{\text{B}\rightarrow \text{A}}$, which approaches unity when model A effectively reproduces the variations in model B. The left, center, and right panels correspond to projections of Mead20+, Schneider19+, and Arico24+ onto the other two models, respectively.

We find that the power spectrum variations induced by parameters in the Mead20+ model are almost entirely captured by both the Schneider19+ and Arico24+ models. In contrast, the Mead20+ model struggles to reproduce variations associated with parameter governing the stellar component in the other two models.
In contrast, the Schneider19+ and Arico24+ models exhibit comparable flexibility, and are able to represent each other’s parameter-induced variations with reasonable accuracy.

It is important to note that while a given model may contain unique modes not captured by others, this does not necessarily imply those modes are physically meaningful or useful for modeling data. Conversely, even if two models exhibit comparable flexibility in reproducing variations across observables, one parameterization may offer a more physical connection between different model ingredients, or make it easier to define physically motivated priors. Also, note that we compare the model flexibility for a specific set of probes ($P_{\rm mm},\,P_{\rm ee},\,P_{\rm mp}$) and the results might change as more probes are added to the data vector.

\section{Simulations}
\label{sec:simulation}
The \textsc{Magneticum} suite of cosmological hydrodynamical simulations \citep{Hirschmann2014, Dolag2016, Steinborn2016, Remus2017, Castro2020,Dolag2025} were performed using the smoothed particle hydrodynamics (SPH) code \textsc{P-GADGET3} \citep{Springel2005}, extend by an improved of the SPH hydrodynamical scheme \citep{Beck2016}. These simulations incorporate a comprehensive subgrid model including radiative cooling and heating in the presence of a uniform ultraviolet background \citep{Springel2005b}, star formation following a multiphase interstellar medium model \citep{Springel2003}, and metal-dependent cooling informed by stellar evolution and chemical enrichment models \citep{Tornatore2007, Wiersma2009}. Feedback from supernovae is modeled via galactic winds, while black hole growth and active galactic nuclei (AGN) feedback are included through a modified version of the model by \citet{Springel2005}, further developed in \citet{Hirschmann2014} to account for both quasar and radio modes of AGN feedback. The \textsc{Magneticum} simulations have been extensively validated against a broad range of observational data across galactic and cluster scales. They successfully reproduce key benchmarks such as the galaxy stellar mass function \citep{Lustig2022}, X-ray scaling relations of galaxy clusters \citep{Biffi2018,2022A&A...666A..22R}, properties of the intra-cluster medium \citep{McDonal2014, Gupta2017,2025A&A...695A.282E}, and the amplitude and shape of the Sunyaev-Zel’dovich power spectrum \citep{Dolag2016}. A full description of the simulation methodology and physical model can be found in Sections 2 \& 3 of \citet{Dolag2025}.

In this work we use the \texttt{Box2b/hr} simulations that have a box size of 640 Mpc$/h$. The simulation is run using a total of $2\times2880^3$ particles with gas and dark matter mass resolutions of $m_{\text{gas}}=1.4\times10^8$ M$_\odot/h$ and $m_{\text{DM}}=6.9\times10^{8}$ M$_\odot/h$, respectively. The simulations are run at a WMAP7 cosmology \citep{Komatsu2011}, the cosmology parameter are: $\Omega_\text{m}=0.272, \Omega_\text{b}=0.0456,  \Omega_\Lambda=0.728, \sigma_8=0.809, h=0.704$, and  $n_\text{s}=0.963$.

\section{Analysis setup and philosophy}
\label{sec:analysis_setup}
We evaluate the performance of analytical models in describing matter clustering and baryonic thermodynamics. The tests are motivated by two main ideas (a) assessing if models can simultaneously and accurately describe the power spectra that source the various cosmological and astrophysical probes, and (b) evaluating whether they connect different components (e.g., dark matter, gas and stars) in a physically meaningful way or simply act as an \textit{effective} model. As power spectra are compressed statistics that integrate contributions over a wide range of scales and halo masses, it is not guaranteed that the inferred properties (e.g., temperature profiles of group- and cluster-sized halos) faithfully reflect the underlying halo population.\footnote{ While \citet{Giri2021} showed that their model could recover baryonic mass fractions when fit to the matter suppression in hydrodynamical simulations, the other two models have not been tested in this way. Therefore, these tests, especially the consistency between predicted thermodynamic profiles and simulations, goes well beyond the original design and validation of these models.} This makes such validation important for ensuring reliable interpretation of multi-probe data.\footnote{We caution that both the accuracy of the power spectrum fits and the agreement of inferred halo properties with simulations depend on how well the halo model describes matter clustering. The shortcomings of the vanilla halo model are well known \citep[e.g., see section 5 in ][]{Asgari2023},  and the specific extensions used to address them can impact the analysis results.}

We jointly fit each model to the following tracer combinations:
\begin{enumerate}
\item matter–matter ($R_{\rm mm}$) and matter–pressure ($R_{\rm mp}$) power spectrum response,
\item  matter–matter ($R_{\rm mm}$) and matter–pressure ($R_{\rm mp}$) power spectrum response along with the mass fractions of bound gas ($f_{\rm BG}$) and stars ($f_{\rm star}$),
\item matter–matter ($R_{\rm mm}$), matter–pressure ($R_{\rm mp}$), and electron density ($R_{\rm ee}$) power spectrum response. 
\end{enumerate}
Using the model fits to data sets 1–3, we predict the stellar and bound gas mass fractions as functions of halo mass, as well as the electron pressure, density, and temperature profiles for group-scale ($13 \leq \log (M_{\rm 200c}h/\mathrm{M}_\odot) < 13.5$) and cluster-scale ($14.5 \leq \log (M_{\rm 200c}h/\mathrm{M}_\odot) < 15$) halos, and compare these predictions to simulation measurements.

\subsection{Power spectrum response}
Following \citetalias{Mead2020}, we adopt a \textit{response}-based approach for fitting the power spectra from hydrodynamical simulations.
Modeling the power spectrum response (Eq.~\refc{eq:response}) has the advantage that in the case of the matter power spectrum, it cancels the fluctuations due to the cosmic variance and also ensures that there is sufficient power in the transition region \citep[e.g.,][]{Tinker2005,Mead2015}. Once the response is calibrated by fitting to a hydrodynamical simulation, it can be combined with any non-linear matter power spectrum --- such as from emulators trained on \textit{N}-body simulations, for e.g., \texttt{Halofit} \citep{Takahashi2012}, \texttt{Bacco} \citep{Angulo2021}, or \texttt{EuclidEmulator2} \citep{EE2} --- to obtain the power spectrum including baryonic effects. For the theory prediction, the response is calculated with respect to the matter power spectrum of a universe that is entirely composed of dark matter halos with an NFW density profile (refer to the introduction of Sec.~\refc{sec:halo_models} for details). For the \textsc{Magneticum} simulation, the response is measured with respect to the phase-matched dark matter-only counterpart to the hydrodynamical simulation.

\citetalias{Mead2020} find that the response-based approach is less effective for power spectra other than matter--matter since the transition regime occurs at different scales. In addition, the one-halo term of the matter-pressure power spectrum is dominated by a small number of massive halos and has a significant contribution even on large-scales. As a result, the errors on large-scale are non-Gaussian and we do not benefit from the same cancellation of cosmic variance that occurs in the matter--matter case. To address this issue we follow the prescription from \citep{Mead2021} and include a smoothing parameter so that the one- and two-halo terms are combined following Eq.~(\refc{eq:Pk_smooth}). The free parameter $\alpha_\text{smooth}$ is allowed to vary in the range $(0.5, 1.5)$. Although this parameter is not physically motivated, the added flexibility is useful for modeling the transition regime and significantly improves the fit quality. We adopt separate smoothing parameters for power spectra of the pressure and electron density fields.  It is also worth highlighting that the \textsc{Magneticum} \texttt{Box2b} simulation used in this work has approximately four times the volume of the \textsc{BAHAMAS} simulation used in \citetalias{Mead2020}. This makes the measurements of the matter--pressure power spectrum significantly less noisy on large-scale, which is important when fitting it with a smooth model prediction.

\subsection{Likelihood}
We fit measurements from the $z=0.25$ snapshot in \textsc{Magneticum} simulation at the WMAP7 cosmology. We  sample model parameter using the \textsc{Nautilus} nested sampler\footnote{\murl{https://nautilus-sampler.readthedocs.io/en/latest}} \citep{nautilus}, assuming a Gaussian likelihood 
\begin{align}
    \log \mathcal{L} = \log \mathcal{L}_{\rm response} + \log \mathcal{L}_{\rm mass\,fraction},
\end{align}
where $\mathcal{L}_{\rm response}$ and $\mathcal{L}_{\rm mass\,fraction}$ denote the contribution of the power spectrum response and mass fractions, respectively. Note that, the second term is included only in cases where the model is fit jointly to the power spectrum and mass fraction data.

The likelihood for the mass fraction component is
\begin{align}
\label{eq:like_mfrac}
    \log \mathcal{L}_{\rm mass\,fraction} = -\frac{1}{2}\sum_X\sum_j \biggl(\frac{f_X^\text{sim}(M_j) - f_X^\text{model}(M_j)}{\sigma_X(M_j)}\biggr)^2,
\end{align}
where the superscript denotes if the quantity is measured from the simulation or computed from the model. The subscript $X=\{{\rm BG}, {\rm star}\}$, so that $f_{\rm BG}(M_j)$ and $f_{\rm star}(M_j)$ represent the mass fractions of the bound gas\footnote{To be consistent with simulation measurements, we define the bound gas fraction as $\frac{\int_0^{R_{\rm 200c}} \rho_{\rm gas}(r)\,{\rm d}r}{\int_0^{R_{\rm 200c}} \rho_{\rm m}(r)\,{\rm d}r}$, where $\rho_{\rm gas}(r)$ and $\rho_{\rm m}(r)$ are the \textit{total} gas and matter density profiles, respectively. This is only important for the Mead20+ and Arico24+ models in which the density profile of the ejected gas component also contributes at $R<R_{\rm 200c}$.} and stars in mass bin $j$, respectively. Similarly, $\sigma^2_{\rm BG}(M_j)$ and $\sigma^2_{\rm star}(M_j)$ is the variance in the bound gas and stellar mass fraction in mass bin $j$. The mass fractions are measured from the simulation in four equally spaced log-bins from $10^{13}-10^{15}\,{\rm M}_\odot/h$ as average over 200 halos each (refer to Appendix~\refc{appx:Prof_magneticum} for details). The model predictions are similarly evaluated at the same halo masses as those selected from the simulation and averaged in each mass bin. We assume that the mass fractions are estimated with five-percent uncertainty in all mass bins i.e., $\sigma_{\rm BG}=\sigma_{\rm star}=5\%$.

The likelihood for the power spectrum response is
\begin{equation}
\label{eq:like_response}
    \log \mathcal{L}_{\rm response} = -\frac{1}{2}\sum_i \biggl(\frac{R_i^\text{sim}(k) - R_i^\text{model}(k)}{\sigma_i(k)}\biggr)^2,
\end{equation}
here, $R_i(k)$ denotes the power spectrum response for probe $i\in\{\text{mm},\,\text{mp},\,\text{ee}\}$ i.e., matter--matter, matter--pressure, and electron density--electron density, respectively. The measurement of the power spectrum response is summarized in Appendix~\refc{appx:Pk_magneticum}. We fit the power spectrum response in 100 linearly spaced bins between  $k\sim0.03-4\,\text{Mpc}^{-1}h$ and assume a diagonal covariance  $\sigma_i(k)=5\%$. The latter choice allows us to asses the accuracy and flexibility of the models independent of survey-specific details (e.g., similar approach was adopted in \citetalias{Arico2024}). 
In practice, the covariance is scale-dependent and non-linear evolution induces coupling between different scales \citep{Krause2017}. Accounting for the full covariance would alter the relative importance of different scales, potentially allowing the model more leeway in fitting the data. Likewise, our current setup treats all probes as contributing equally to the total likelihood, whereas in realistic analyses, the cross-covariance between the various probes might also make it easier for the model to simultaneously fit them. Although, since the sensitivity to feedback processes depends on halo mass, this reweighting (of different scales and probes) can affect parameter constraints and shift the inferred halo properties. We defer the analysis with a survey-realistic covariance including the non-Gaussian effects to a future study.

\subsection{Priors}
The full list of varied parameters and priors for each model is provided in Table~\refc{tab:model_params}. We use wide uniform priors for most parameters with the range derived from previous applications of these models to hydrodynamical simulations \citep{Schneider2015,Schneider2019, Arico2020, Arico2021, Mead2020, Pandey2024}. However, certain model ingredients, such as the satellite parameters in BCM, remain poorly constrained due to limited simulation studies. As a result, the priors adopted for these parameters may not be optimal and can potentially be improved in future work. Our choice of the prior for the smoothing parameter, $\alpha_{\rm smooth}\in\mathcal{U}(0.5,1.5)$, is informed by preliminary analysis where the marginalized posterior estimates typically spanned 0.6 to 1.3. We adopt a unified, slightly broader range to ensure that our results are not prior-dominated.

In preliminary tests, we found that even with physically motivated priors some models produced unphysical results, for e.g. stellar fractions exceeding the total gas fraction at certain halo masses. To avoid such cases, we impose constraints to exclude these unrealistic regions of parameter space:
\begin{enumerate}[label=\roman*]
    \item $f_{\rm gas}(M)>f_{\rm star}(M)$,  $\forall M\in(10^{12},10^{15})\,{\rm M}_\odot h^{-1}$,
    \item $f_{\rm cga}(M)>f_{\rm sga}(M),\, M\simeq10^{12}\,{\rm M}_\odot h^{-1}$,
    \item $f_{\rm cga}(M)<f_{\rm sga}(M),\, M\simeq10^{15}\,{\rm M}_\odot h^{-1}$,
    \item $f_{\rm star}(M)>0.5\%,\, M\simeq10^{15}\,{\rm M}_\odot h^{-1}$
    \item  $f_{\rm gas}(M),f_{\rm star}(M)>0,\,\forall M\gtrsim10^{10}\,{\rm M}_\odot h^{-1}$.
\end{enumerate}
(i) makes sure that the baryonic budget is dominated by gas. Constraints (ii) and (iii) enforce physically motivated trends in the galaxy population: central galaxies dominate the stellar content at low masses, while satellites dominate at high masses \citep{Kravtsov2018}.  Condition (iv) ensures that high mass halos have at least some stars. As described in the introductory text for Sec.~\refc{sec:halo_models}, the signal from halos below  $M\sim10^{10}\,{\rm M}_\odot h^{-1}$ is accounted for by using an analytic correction and (v) is necessary for this correction to be non-zero when predicting the statistics of the gas fields.

Additionally, we impose a specific constraint for the Arico24+ model. The inner slope of the gas density profile, $\beta_{\rm i}$ (Eq.~\refc{eq:AA_betai}), can become negative for certain halo masses, leading to unrealistic features in the gas density profile: for $r \ll \theta_{\rm out} r_{\rm 200c}$, $\frac{{\rm d}\rho_{\rm BG}}{{\rm d}r} > 0$ i.e., the gas density would increase with radius. To avoid such scenarios, we require $\beta_{\rm i} > 0$ for all $M \gtrsim 10^{10}\,{\rm M}_\odot h^{-1}$.

In models such as the ones considered here, we have a wide array of input parameters controlling the different halo components across a range of masses and redshifts. As the dimensionality increases, so does the volume of parameter space where the model produces unphysical predictions. Previous approaches have handled this by fine-tuning the priors or forcing the model to be physical (e.g., by altering the stellar fraction when the combined gas and star fraction exceeds the baryon fraction). The use of constraint equations is a simpler approach for enforcing prior knowledge from astrophysics and structure formation.

\begin{table*}
\centering
\begin{tabular}{ll cc |cccc |ccc}
\toprule
 & \textbf{Model}& \multicolumn{9}{c}{\textbf{RMS Error} [\%]}  \\
 & & \multicolumn{2}{c}{(1)}  & \multicolumn{4}{c}{(2)}  & \multicolumn{3}{c}{(3)} \\

\cmidrule(lr){3-4} \cmidrule(lr){5-8} \cmidrule(l){9-11}
& & $R_{\rm mm}$ & $R_{\rm mp}$ & $R_{\rm mm}$ & $R_{\rm mp}$ & $f_{\rm BG}$ & $f_{\rm star}$ & $R_{\rm mm}$ & $R_{\rm mp}$ & $R_{\rm ee}$ \\

\cmidrule(r){1-2} \cmidrule(lr){3-4} \cmidrule(lr){5-8} \cmidrule(l){9-11}

\multirow{3}{*}{\textbf{Six-parameter set}} & Mead20+    & 0.2& 0.8 & 0.9 & 1.0 & 7.7& 41.1 & 1.6 & 4.6 & 4.3\\
& Schneider19+  & 0.3& 0.7& 1.2 & 0.9 & 9.1 & 19.2& 1.6 & 1.7 & 1.0 \\
& Arico24+    &  1.8& 0.7 & 1.8 & 0.6 & 1.5& 29.5 & 2.4 & 3.4 & 4.2 \\ \hline
\multirow{3}{*}{\textbf{Extended set}} & Mead20+ & 0.3& 0.8& 0.4 & 0.9 & 2.0 & 4.7 & 1.0 & 1.3 & 0.9\\
& Schneider19+  & 0.2& 0.5& 0.3& 0.8 & 3.8 & 1.2 & 0.6 & 1.1 & 0.6\\
& Arico24+    & 0.3& 0.6& 0.7 & 0.9 & 1.1 & 1.5 & 0.4 & 1.0 & 0.5    \\
\bottomrule
\end{tabular}
\caption{Model accuracy for joint fits to power spectra and mass fractions  from the \textsc{Magneticum} simulation at $z=0.25$. We fit to combinations of matter--matter ($R_{\rm mm}$), matter--pressure ($R_{\rm mp}$), and electron density ($R_{\rm ee}$) power spectrum response as well as the bound gas ($f_{\rm BG}$) and stellar mass fraction ($f_{\rm star}$). Columns (1), (2), and (3) shows the root mean squared (RMS) error of the best fit model prediction and simulation measurement for each combination. For each model, we perform the fits with a six-parameter model and a variant with additional free parameters referred to as the extended set. Parameters for both sets are listed in Table \protect\refc{tab:model_params}.}
\label{tab:results_mm_mp}
\end{table*}

\begin{figure*}
    \centering
    \includegraphics[width=\linewidth]{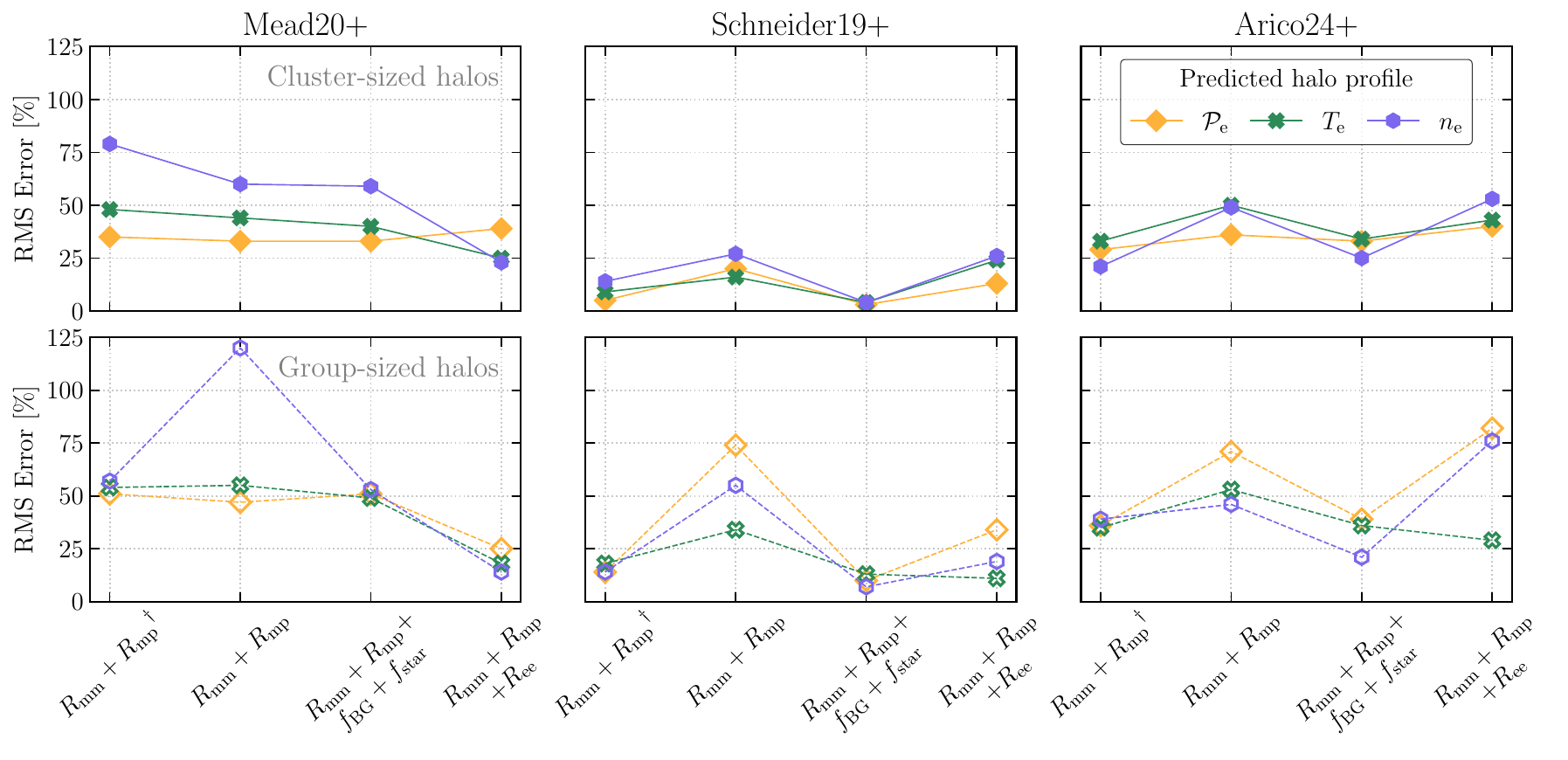}
    \caption{Accuracy of predicted thermodynamic profiles from models calibrated on power spectra and mass fractions. Columns from left to right show results for the Mead20+, Schneider19+, and Arico24+ models, respectively. The horizontal axis shows the tracer combination on which the model is calibrated. Tracer combinations marked by $\dagger$ are fit using a six-parameter version of the models while the remaining combination are analyzed with the extended set (Table \protect\refc{tab:model_params}). The vertical axis shows the RMS error on the electron pressure ($\mathcal{P}_{\rm e}$), temperature ($T_{\rm e}$), and density ($n_{\rm e}$) profiles. We compute the RMS error between the best-fit prediction of the profile and simulation measurement, over the range $0.12<r/r_{\rm 200c}\leq1$ for cluster-sized halos (top row) and $0.25<r/r_{\rm 200c}\leq1$ for group-sized halos (bottom row).}
    \label{fig:profile_rms}
\end{figure*}

\section{Results}
\label{sec:results}
In this section we present results from jointly analyzing the different tracer combinations with the three models. In Sec.~\refc{sec:res_min} we jointly fit $R_{\rm mm}$ + $R_{\rm mp}$ responses  while in Sec.~\refc{sec:mm_mp_ext} we fit the same combination with an extended parameter set. In Sec.~\refc{sec:mm_mp_mfrac} we include mass fractions ($f_{\rm BG}$ and $f_{\rm star}$)  as additional inputs. In Sec.~\refc{sec:mm_mp_nene} we instead use $R_{\rm ee}$ as the additional input. As will be described later, the parameter posteriors along with the best-fit values for all these cases are shown in Appendix~\refc{appx:param_posteriors}, while Table~\refc{tab:results_mm_mp} reports the accuracy of the model fits for the respective tracer combinations. We also evaluate how well the baryonic properties inferred from the fits to above tracer combinations match direct measurements from the simulation. The results of this exercise are summarized in Fig.~\refc{fig:profile_rms} and will be discussed later in the subsequent sections.

\subsection{Matter and electron pressure}
\subsubsection{Six-parameter set}
\label{sec:res_min}

We first jointly fit the matter--matter and matter--pressure response using a minimal version of each model which comprises only six free parameters, these are:
\begin{itemize}
    \item Mead20+: $\epsilon_1,\, \Gamma,\, \log M_0,\, \alpha,\,\log T_{\rm w}, \, \alpha_{\rm smooth,mp}$
    \item Schneider19+: $\delta,\, \mu_\beta,\, \theta_{\rm ej,0},\, \eta,\, \alpha_{\rm nt},\, \alpha_{\rm smooth,mp}$
    \item Arico24+: $\theta_{\rm inn},\, \log M_{\rm c},\, \, \log M_{\rm 1,0},\, A_{\rm th},\,\log T_{\rm w}, \,\alpha_{\rm smooth,mp}$
\end{itemize}
where $\alpha_{\rm smooth,mp}$ modifies the matter--pressure power spectrum in the transition regime according to Eq.~(\refc{eq:Pk_smooth}).

The choice of the six-parameter set for Mead20+ follows \citetalias{Mead2020}, where the authors identified the most important parameters through trial and error. The parameters for the Arico24+ model are chosen based on  \citetalias{Arico2024} (although we fix $\eta$ so that there are a total of six parameters). For Schneider19+, we similarly vary three parameters for the gas component \{$\delta,\, \mu_\beta,\, \theta_{\rm ej,0}$\}, one parameter of the stellar component ($\eta$), and one parameter for non-thermal pressure ($\alpha_{\rm nt}$).

\begin{figure*}
    \centering
    \includegraphics[width=0.92\linewidth]{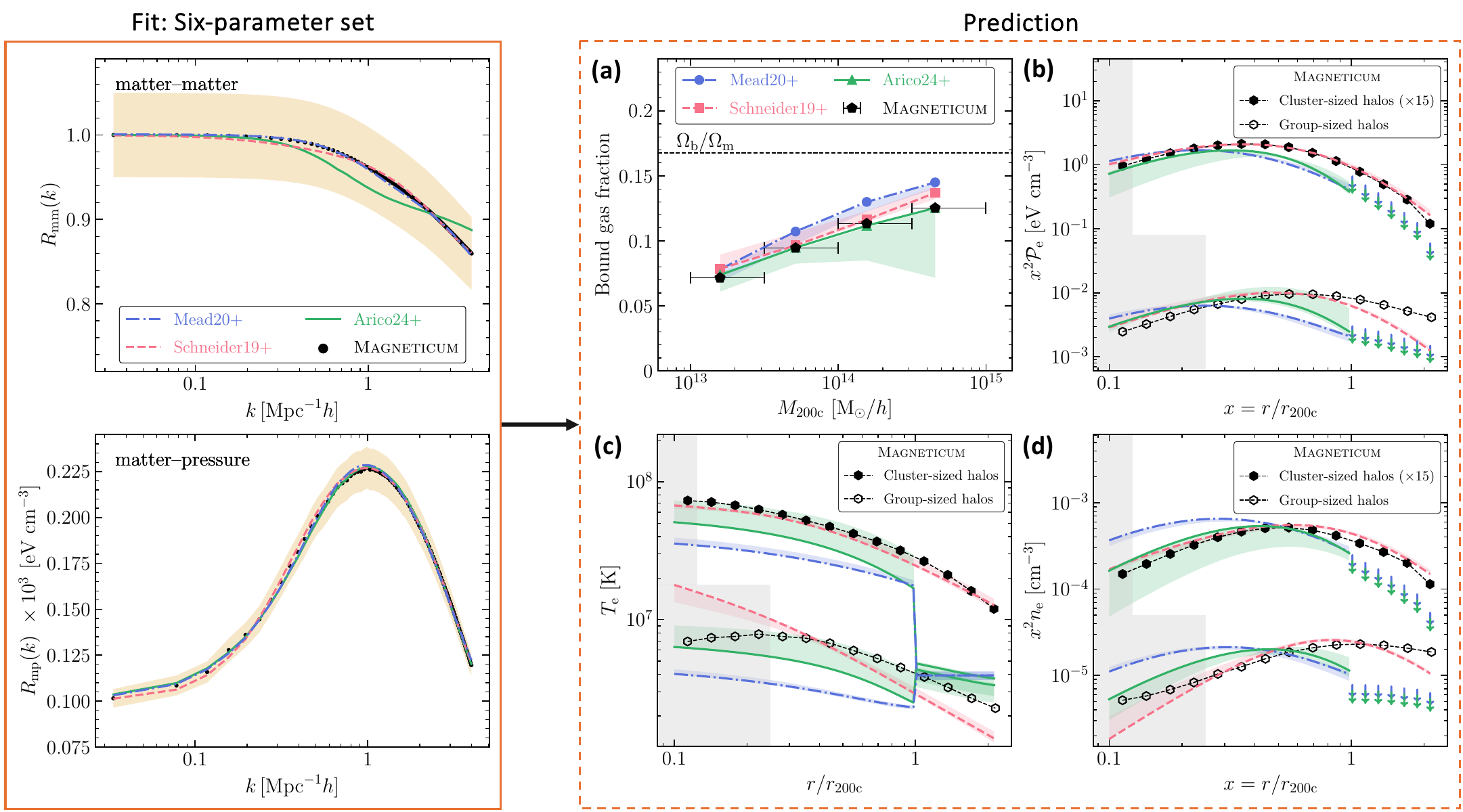}
    \caption{Results from joint fits to $R_{\rm mm}(k)$ and $R_{\rm mp}(k)$ from \textsc{Magneticum} (at $z=0.25$) using a six parameter model (parameters marked with \textcolor{Mahogany}{$\dagger$} in Table.~\protect\refc{tab:model_params}). \textbf{Left:} Best-fit predictions from the three models considered here and simulation measurements. The light shaded region represents the assumed standard deviation of  5\%. \textbf{Right:} Panels (a) through (d) show the predicted bound gas fraction and electron pressure, temperature, and density profiles, respectively, and the colored shaded regions represent the $16^{\rm th}-84^{\rm th}$ percentile regions. While we show full the inferred profiles down to $x=(R/R_{\rm 200c})=0.1$, note that the power spectra responses up to $k<4\,{\rm Mpc}^{-1}h$ that we fit to are only sensitive to physical scales of $x\gtrsim0.25$ for group mass halos ($13\leq\log (M_{\rm200c}h/\mathrm{M}_\odot) <13.5$)  and $x\gtrsim0.12$ for cluster mass halos ($14.5\leq\log (M_{\rm200c}h/\mathrm{M}_\odot)<15$). We use downward arrows to indicate where the model prediction differs from the simulation by $\gtrsim10\times$. See Sec. \protect\refc{sec:res_min} for related discussion.}
    \label{fig:fit_min}
\end{figure*}

The left panel in Fig.~\refc{fig:fit_min} shows the best-fit power spectrum responses from the three models alongside simulation measurements. Table~\refc{tab:results_mm_mp} summarizes the fit accuracy as the root mean square (RMS) error of best-fit model predictions w.r.t. simulation data. We find that, with only six free parameters, all three models can achieve sub-percent accuracy for matter--matter and matter--pressure power spectra, with the exception of the Arico24+ model which achieves percent level accuracy for the matter--matter response. 

The results for the Arico24+ as well as Mead20+ models represent a clear improvement over those of \citetalias{Arico2024} and \citetalias{Mead2020} respectively, who reported $\sim$10-15\% accuracy in modeling the matter--pressure power spectrum. This is primarily due to updated halo model ingredients (see Appendix~\refc{appx:models}) for the two models. Furthermore, the \textsc{Magneticum} \texttt{Box2b/hr} volume is roughly four times larger than the BAHAMAS suite used in the original studies. This increased volume reduces cosmic variance for $P_{\rm mp}$, which is dominated by massive halos, thereby enabling a more precise model fit to the simulation data.  We find that the addition of $T_{\rm w}$ parameter in Arico24+ model provides crucial capability to modulate the electron pressure in the halo outskirts. Moreover, the inclusion of the smoothing parameter ($\alpha_{\rm smooth,mp}$) for the transition regime provides large improvement in the fits quality for both Arico24+ and Mead20+ models. As shown in Fig.~\refc{fig:post_M20}, the marginalized posterior for $\alpha_{\rm smooth,mp}$ significantly deviates from unity for both models.\footnote{On repeating the analysis without the smoothing parameter, the fit quality for the matter--pressure power spectrum degrades to the level reported in the original works.} We further discuss the implications of including transition smoothing in Sec.~\refc{sec:discussion_flex}.

Panel (a) through (d) in Fig.~\refc{fig:fit_min} shows the constraints on inferred halo properties as well as the 16$^{\rm th}$-84$^{\rm th}$ percentile region computed from 200 random samples from the chain. Panel (a) shows the constraints on bound gas fractions (mass of gas relative to total matter within $R_{\rm 200c}$) across different halo masses compare to simulation measurements. All models yield predictions in good agreement with the \textsc{Magneticum} simulation with the Mead20+ and Schneider19+ models somewhat overestimating the gas content for the most massive mas bin. This finding is consistent with previous studies such as \citet{Giri2021} which showed that it is possible to recover mass fractions by fitting to the power suppression. The inferred stellar fractions (presented in Appendix~\refc{appx:fstar_constraint}) are inconsistent with the simulation measurements, especially at the high-mass end.

\begin{figure*}
    \centering
    \includegraphics[width=0.92\linewidth]{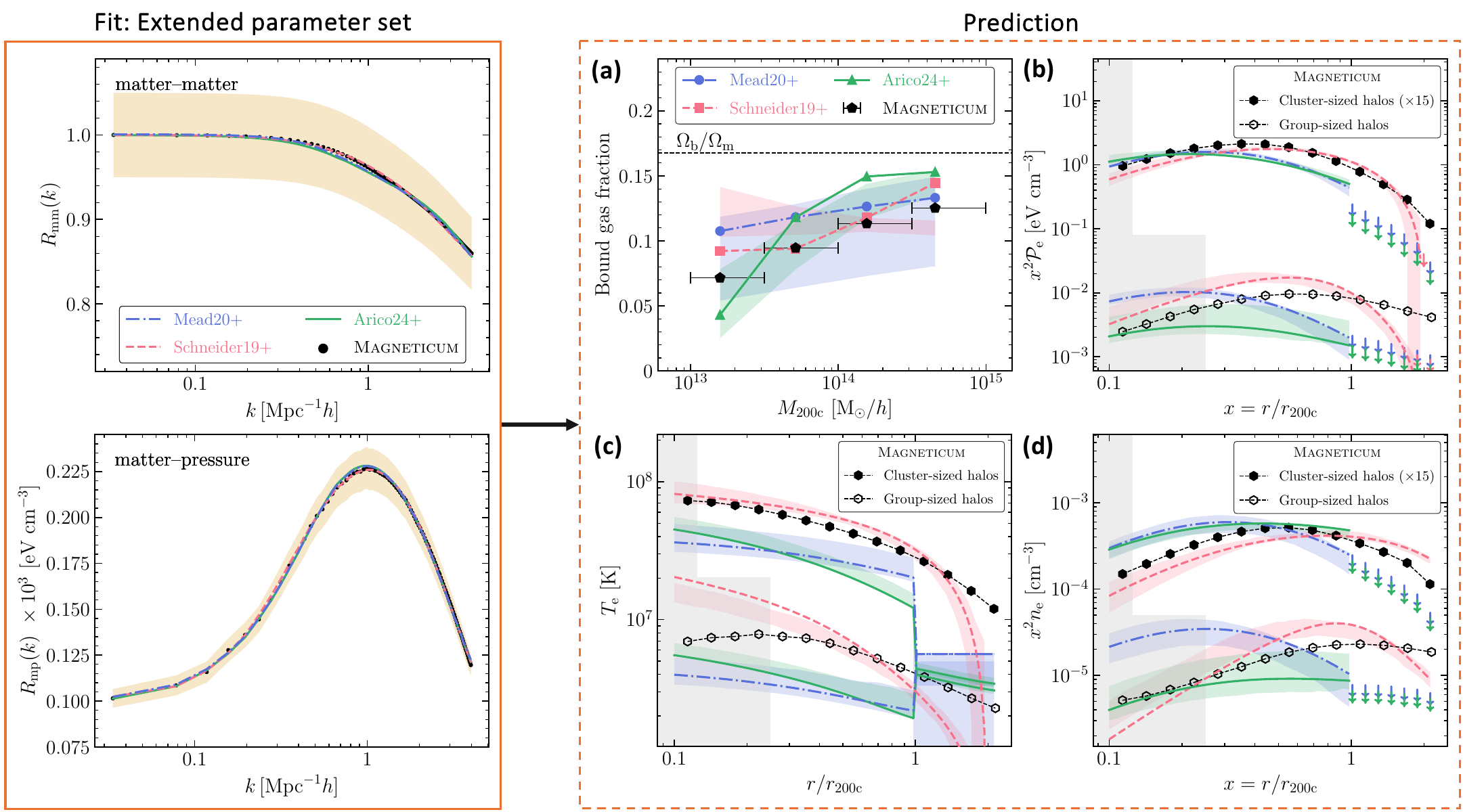}
    \caption{Same as Fig. \protect\refc{fig:fit_min} but fitting $R_{\rm mm}(k)$ and $R_{\rm mp}(k)$ from the \textsc{Magneticum} simulation with an extended parameter set (Table~\protect\refc{tab:model_params}). See Sec. \protect\refc{sec:mm_mp_ext} for related discussion.}
    \label{fig:fit_ext}
\end{figure*}

Panels (b) through (d) in Fig.~\refc{fig:fit_min} show thermodynamic profiles for group-sized ($13\leq\log\frac{M_{\rm200c}}{h^{-1}\mathrm{M}\odot}<13.5$)  and cluster-sized ($14.5\leq\log\frac{M_{\rm200c}}{h^{-1}\mathrm{M}_\odot}<15$) halos. 
The accuracy of the inferred profiles are quantified in Fig.~\refc{fig:profile_rms}. The models reproduce the profiles within the 20-40\%, with the Mead20+ model being most discrepant, especially when predicting the electron density profile.

It is difficult to draw conclusions regarding the accuracy of the power spectra and inferred halo properties, since they are sensitive to the choice of free parameters and values of the fixed parameters. The free parameters were selected based on their ability to reproduce matter power suppression trends in other hydrodynamical simulations. However, when applying the models to an independent simulation like \textsc{Magneticum}, the same set of parameters may not fully capture the mass and scale dependence of feedback processes. The other potential issue concerns the fixed parameters: fiducial values which happen to align with those preferred by \textsc{Magneticum} might enable the model to perform well, otherwise, it might underperform or compensate by shifting the free parameters, potentially biasing the inferred halo properties.

\subsubsection{Extended parameter set}
\label{sec:mm_mp_ext}

We now explore how their accuracy improves when additional parameters are varied. The extended parameter set for the Mead20+, Schneider19+, and Arico24+ models contain 11, 14, and 15 parameters, respectively (see Table~\refc{tab:model_params}) which provide greater flexibility in jointly describing baryon thermodynamics and matter distribution in a large dynamical range of halo masses. We summarize the model accuracy when jointly fitting the matter–matter and matter–pressure power spectrum response in Table~\refc{tab:results_mm_mp} and Figure~\refc{fig:fit_ext}. We see that with this extended set of parameters, the Arico24+ model is also able to achieve a sub-percent accuracy in fitting the responses, similar to Mead20+ and Schneider19+ models.

The right panel of Fig.~\refc{fig:fit_ext} presents the inferred mass fraction and profiles along with the associated 16$^{\rm th}$-84$^{\rm th}$ percentile regions. As expected, increasing the number of parameters increases the uncertainty on inferred quantities, which is also reflected in the broadening of parameter posteriors as shown in Appendix~\refc{appx:param_posteriors}. The inferred bound gas fractions from all models are slightly worse than for the six-parameter set, even though the six-parameter set is a proper subspace of the extended model. However, the mass fractions remain broadly consistent with the simulation within uncertainty. We also find no significant improvement in the stellar fraction even after freeing up more parameters (Fig.~\refc{fig:result_fstar}). The likely reason is that  our power spectrum measurements, which are sensitive to minimum scales of $r/r_{\rm 200c}\sim0.1$, receive a relatively small contribution from stars, which only dominate the profile close to halo centers ($r/r_{\rm 200c}\sim10^{-2}$).

Evaluating the performance of the inferred profiles we find that all models perform worse compared to the six-parameter set, especially for group-sized halos where the difference in the pressure profile reaches up to 75\% for the Schneider19+ model (also see Fig.~\refc{fig:profile_rms}). For cluster-sized halos, which dominate the matter--pressure power spectrum, the pressure profile from Schneider19+ model shows broad agreement while the Mead20+ and Arico24+ models significantly underpredict the amplitude from $r/r_{\rm 200c}\sim0.4$ to all the way to the halo outskirts. Despite this mismatch, both models still fit the power spectrum well, likely due to compensation by the smoothing parameter ($\alpha_{\rm smooth,mp}$), which deviates significantly from unity (see Figs.~\refc{fig:post_M20} and Fig.~\refc{fig:post_A24}) and boosts power near the one-halo to two-halo transition. For Mead20+ and Arico24+, the discrepancy in pressure beyond the halo boundary can also stem from the treatment of ejected gas, which is assumed to be heated to a constant temperature $T_{\rm w}$, independent of radius or halo mass.\footnote{In Arico24, a mass-dependent correction for non-thermal pressure introduces a mild mass dependence to the ejected gas temperature.}

These results raise a key question: why does increasing the number of free parameters not improve, and in some cases worsen, the recovery of inferred halo properties? Parameter degeneracy (see Fig.~\refc{fig:corr_matrix})) and projection effect are major contributing factors when using the parameter posterior to predict the thermodynamic profiles. If the data are not constraining enough or in the absence of informative priors, the inferred profiles can drift away from the physically correct solution. Input summary statistics represent an aggregate effect of all halo masses, and overfitting or lack of constraining power can bias the inference of the different baryonic profiles for individual halo mass bins. Although we applied strategies like re-binning the power spectra, which smooths the large-scale noise in the matter–pressure spectrum, models with $\mathcal{O}(15)$ free parameters may still be too flexible for this particular data set comprising two power spectrum responses. Therefore, as we show in the following subsection, if the model is physically accurate, we would expect that including more summary statistics probing complementary baryonic physics would alleviate these issues.

\subsection{Matter, electron pressure, and mass fractions}
\label{sec:mm_mp_mfrac}

We now extend the analysis by incorporating the bound gas and stellar mass fractions directly into the fits. This allows us to test whether the models can simultaneously reproduce both the power spectra and mass fractions, and to assess whether including this additional information improves the inference on halo properties.

\begin{figure*}
    \centering
    \includegraphics[width=\linewidth]{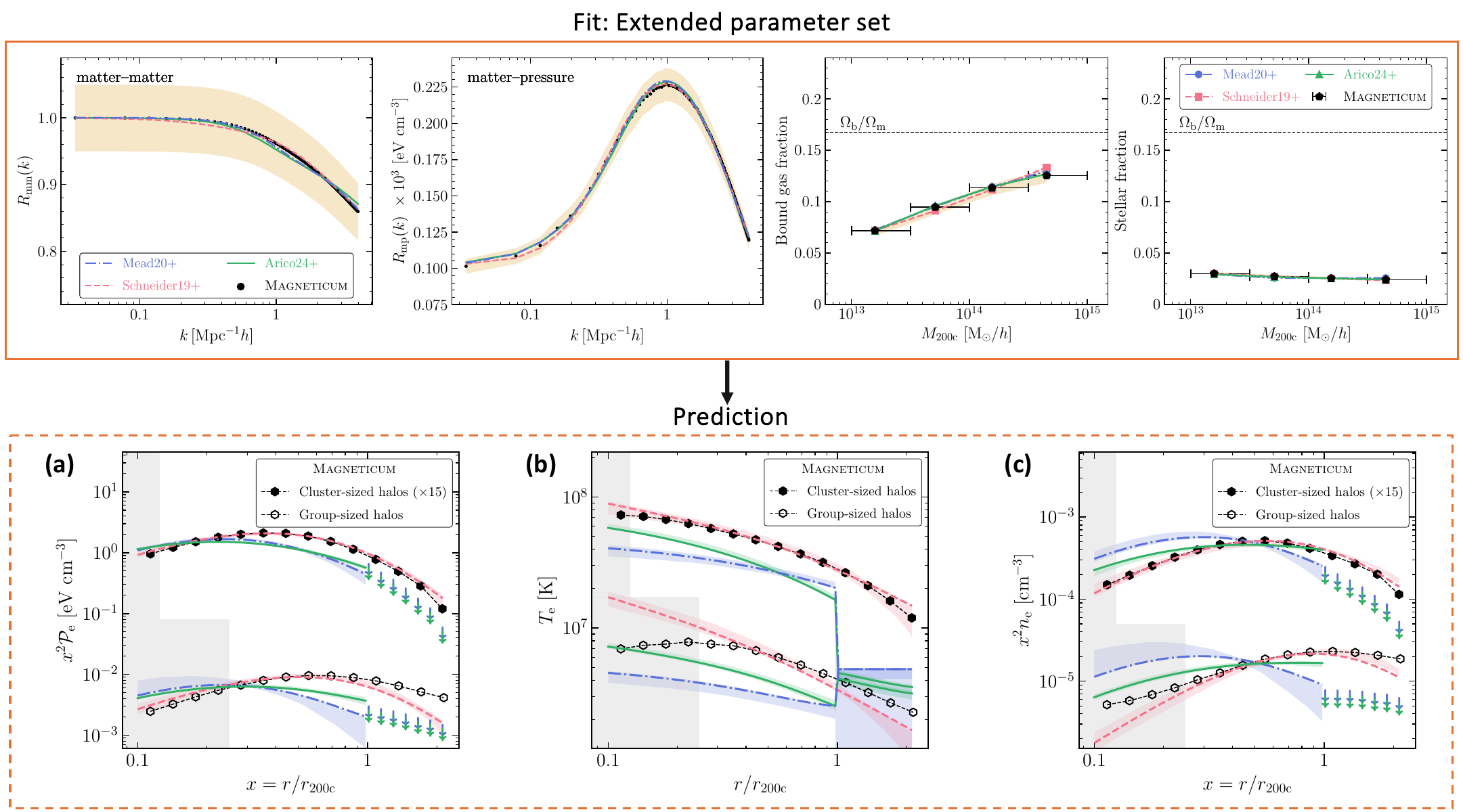}
    \caption{Same as Fig. \protect\refc{fig:fit_min} but fitting $R_{\rm mm}(k)$, $R_{\rm mp}(k)$, $f_{\rm BG}$, and $f_{\rm star}$ with an extended parameter set (Table~\protect\refc{tab:model_params}) and predicting electron pressure, temperature, and density profiles. See Sec. \protect\refc{sec:mm_mp_mfrac} for details.}
    \label{fig:fit_mm_mp_mfrac}
\end{figure*}

The results of fitting the six-parameter model are summarized in Table~\refc{tab:results_mm_mp}. Compared to fitting matter and electron pressure (Sec.~\refc{sec:res_min}), the accuracy of fit to power spectrum responses is slightly reduced. All models fit the bound gas fraction fairly well, with the Mead20+ and Schneider19+ models slightly over and underestimating the amount of gas, respectively. Neither model adequately captures the stellar mass fraction, particularly in high-mass halos, which is not surprising since we vary at most one parameter for the stellar component.

For the extended set of parameters, we find that all three models demonstrate sufficient flexibility to simultaneously capture the power spectrum responses as well as the gas and stellar mass fractions of halos as shown in Fig.~\refc{fig:fit_mm_mp_mfrac} and quantified in Table~\refc{tab:results_mm_mp}. The power spectrum responses are recovered at sub-percent level accuracy, and the inferred mass fractions are matched with an RMS error of $\approx5$\%. The addition of mass fraction data also improves constraints on model parameters. As seen in Figures~\refc{fig:post_M20},~\refc{fig:post_S19}, and~\refc{fig:post_A24}, the contours for the parameters governing the fractions of gas and stars shrink. Moreover, in Table~\refc{tab:bf_params}, the best-fit values of $\beta$ (slope of bound gas to halo mass relation) for the Arico24+ and Mead20+ models are also close to the values they are fixed to in the six-parameter set. This explains why the gas mass fractions in Fig.~\refc{fig:fit_min} are in such good agreement with simulation measurements, since the fiducial value of $\beta$ is close to what is preferred by \textsc{Magneticum} and therefore we can pin down the \textit{half-way mass} ($M_{\rm c}$ and $M_0$ for Arico24+ and Mead20+, respectively).

Turning to the inferred profiles, we find that including the mass fractions does not result in significant improvement for the Mead20+ model, while the other two show agreement within $40$\% (also see Fig.~\refc{fig:profile_rms}). Specifically, the Schneider19+ model matches the profiles quite well, achieving RMS error $\lesssim10$\% at both group- and cluster-scales.

In summary, on including the mass fractions of gas and stars we find that:
\begin{itemize}
    \item The Schneider19+ model provides an excellent match to simulation profiles at cluster scales but performs less well for group-sized halos. This discrepancy may arise from the mass-independent treatment of inherently mass-dependent quantities (e.g., non-thermal pressure) which may not extrapolate reliably to lower masses, especially since the parameter fits are dominated by the more massive halos.
    
    \item In comparison, the Mead20+ and Arico24+ models yield poorer predictions for the halo profiles. This is somewhat unexpected, as matching the suppression in the matter power spectrum and gas abundance should, in principle, constrain the spatial distribution of gas.

    \item This suggests a potential issue in how the pressure/gas component is modeled, such as: the breakdown of the constant polytropic equation of state near the halo boundary \citep{Battaglia:2012:ApJ:}, or the assumption that ejected gas is heated to the same temperature across all halo masses.
    
    \item For Mead20+ the discrepancy could also be a result of how gas is linked to the dark matter distribution. The model assumes that gas and dark matter share the same modified concentration (Eq.~\refc{eq:HMx_cmod}), but simulations and recent observations show that, unlike dark matter, gas concentration increases with halo mass \citep{Popesso2024} --- indicating that the Mead20+ prescription might simply be unphysical.
\end{itemize}

\subsection{Matter, electron pressure, and electron density}
\label{sec:mm_mp_nene}

\begin{figure*}
    \centering
    \includegraphics[width=\linewidth]{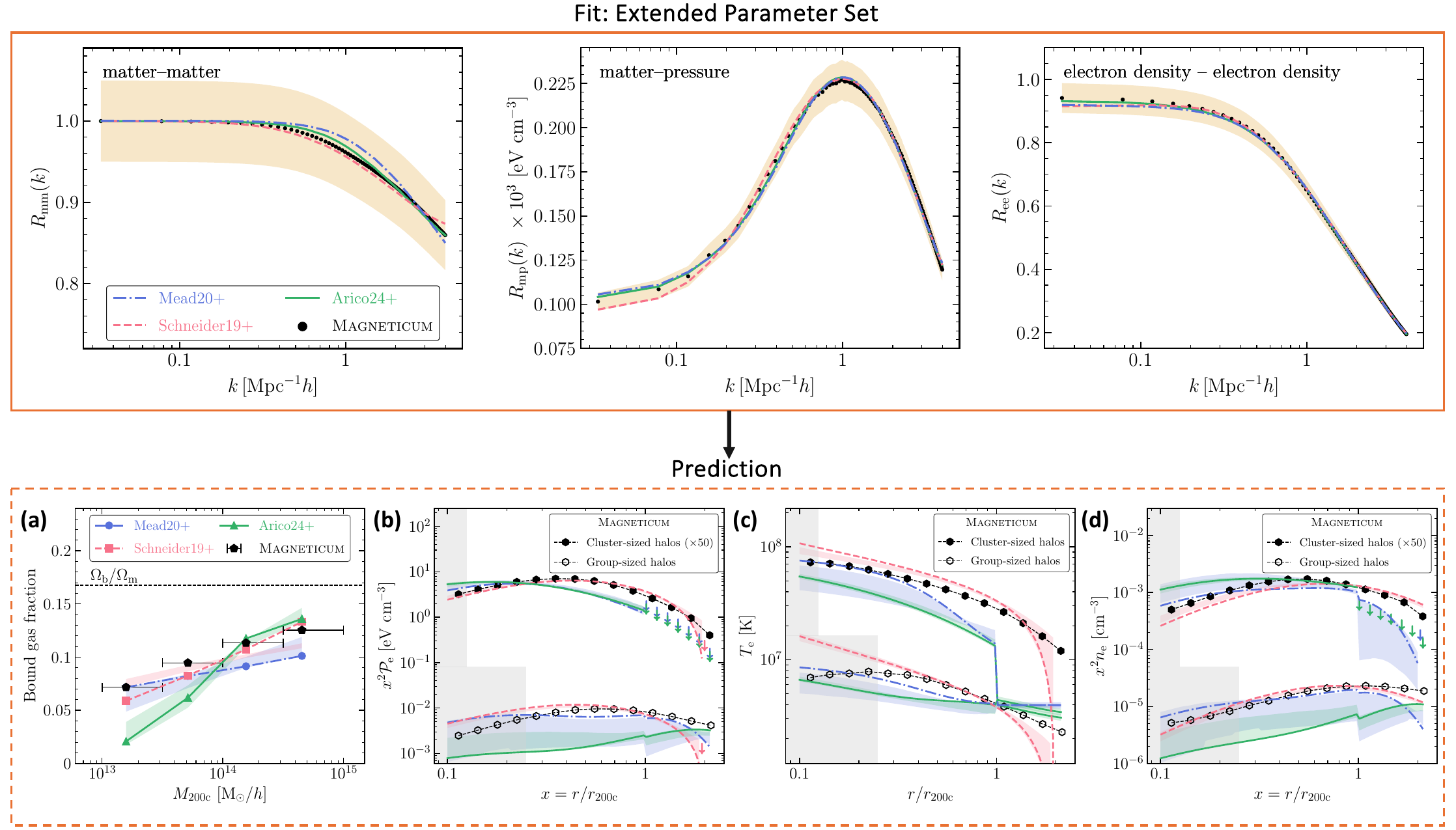}
    \caption{Same as Fig. \protect\refc{fig:fit_min} but fitting $R_{\rm mm}(k)$, $R_{\rm mp}(k)$ and $R_{\rm ee}(k)$ with an extended parameter set (Table~\protect\refc{tab:model_params}) and predicting bound gas fraction along with electron pressure, temperature, and density profiles. See Sec. \protect\refc{sec:mm_mp_nene} for details.}
    \label{fig:Rk_fit_nene}
\end{figure*}

We now test whether the models can simultaneously describe the matter, pressure, and electron density power spectra which source the weak lensing, tSZ, and kSZ (and FRB dispersion measure) observables, respectively.

The results for the six-parameter fit are summarized in column (3) of Table~\refc{tab:results_mm_mp} (figure not shown). All models reproduce the power spectra at the five-percent level. The overall fit accuracy for the matter--matter and matter--pressure responses degrades compared to fits where we only consider the two power spectra. This suggests that the six-parameter set might not offer sufficient flexibility for jointly fitting the three power spectra, especially since the electron density power spectrum is more sensitive to the details of the gas distribution which is challenging to capture with a limited number of parameters.

Figure~\refc{fig:Rk_fit_nene} shows the results from repeating the fits to the power spectra responses with the extended parameter set. As summarized in column (3) of Table~\refc{tab:results_mm_mp}, all models achieve percent-level accuracy in fitting all three input power spectra responses. 

These results are noteworthy because  \citetalias{Mead2020} found it  especially difficult to simultaneously model the electron pressure and gas distribution in halos. Their analysis finds a mean error (averaged over $k=0.015-7\,{\rm Mpc}^{-1}h$) of $\approx30\%$ when the gas power spectrum is predicted from joint fits of the matter--matter and matter--pressure power spectrum. As discussed earlier, the primary improvement in our analyses comes from smoothing parameter employed for modeling the transition region between one- and two-halo terms, which we find is quite important for the matter--pressure power spectrum when using Mead20+ and Arico24+ models.

Next, we compare the inferred halo properties to simulation measurements. Panel (a) in Fig.~\refc{fig:Rk_fit_nene} presents the inferred constraints on the bound gas fraction. For S19+, the inferred bound gas fractions and the stellar fractions (shown in Fig.~\refc{fig:result_fstar}) are in good agreement with the simulations. The Arico24+ model also predicts the correct bound gas fraction for the two highest mass bins but severely underestimates the gas content of group-scale halos. This is also seen in Fig.~\refc{fig:post_A24} where $\beta$ and $M_{\rm c}$ now shift to a larger value, so that the bound gas fraction to halo mass relation becomes steeper and gas in lower mass halos is expelled more efficiently. The Mead20+ model exhibits a deficit of bound gas, particularly in the most massive halos.

Comparing the inferred halo profiles in panels (b) through (d), we find that the Mead20+ model shows the most improvement, reproducing profiles with $15-25$\% agreement, while the other two models show poorer performance compared to the previous case where we included mass fractions in the fit (also see Fig.~\refc{fig:profile_rms}). The Mead20+ and Schneider19+ models match the profiles at group-scales better than at cluster-scales, which is expected since including the electron density power spectrum in the fit shifts the sensitivity to lower mass halos. The Arico24+ model on the other hand, performs significantly worse for group-sized halos, with RMS error of $\approx80$\% for the pressure profile.

In Appendix~\refc{appx:fit_all}, we perform a joint fit using all available tracers: the three power spectrum responses ($R_{\rm mm}, R_{\rm mp}, R_{\rm ee}$), the bound gas and stellar mass fractions, and the thermodynamic profiles. The results, presented in Fig.~\refc{fig:res_mm_mp_nene_mfrac_prof}, show that although the Arico24+ and Mead20+ models reproduce the profiles well, their fit to the power spectra becomes worse. This highlights the limited flexibility of these models and helps explain why the profiles inferred from them are less accurate.

In summary, on including the electron density power spectrum we find:
\begin{itemize}
    \item For Mead20+, the inferred halo profiles at both group- and cluster-scales improve compared to previous cases and fit the simulations at 25-40\%. However, the model underpredicts the amount of bound gas in cluster-mass halos. For Arico24+ model, the inferred profiles show RMS error of 40-80\%. 
    
    \item The Schneider19+ predicted profiles degrade relative to the case when mass fractions are included, but remain broadly consistent with RMS error of 10--30\%. 

\end{itemize}

\section{Discussion}
\label{sec:discussion}
\subsection{Summary}
We have presented a comparison study of three halo models from the literature: Mead20+, Schneider19+, and Arico24+, evaluating their performance in joint analyses of power spectra of cosmological fields (matter–matter, matter–pressure, and electron density-electron density) which source cosmological probes such as weak lensing, the thermal and kinematic Sunyaev-Zel’dovich effects and dispersion measure of fast radio bursts observations.

These models adopt different assumptions for describing the various components of the matter distribution, so understanding their strengths and limitations is essential for reliably extracting physical insight from multi-probe analyses. We show that the model parameters are highly degenerate for predictions of the matter power spectrum, and that adding information about baryons at different density and temperatures (through measurements of the power spectrum or scaling relations) helps break these degeneracies (Fig.~\refc{fig:corr_matrix}). Importantly, the extent of parameter decorrelation varies between models, emphasizing that the specific modeling choices are at least as important as overall model flexibility.

We fit these models to different combinations of measurements (matter–matter, matter–pressure, and electron-electron power spectra responses, bound gas and stellar mass fractions) from the \textsc{Magneticum} simulation, which are independent of the data used to build or calibrate these models. Our main findings are: 

\begin{itemize}
    \item Using a six-parameter set, all models can jointly fit all the power spectra within a few percent (see Table.~\refc{tab:results_mm_mp}).
    
    \item Using $\mathcal{O}(15)$ free parameters, all models jointly fit the power spectra responses with percent-level accuracy (see Table.~\refc{tab:results_mm_mp}).

    \item We find that the Schneider19+ model performs best overall at inferring the thermodynamic profiles at group-scale ($13\leq\log (M_{\rm200c}h/\mathrm{M}_\odot) <13.5$) and cluster-scale ($14.5\leq\log (M_{\rm200c}h/\mathrm{M}_\odot) <15$) halos across most of the probe combinations considered here (see Fig.~\refc{fig:profile_rms}). Particularly, this model achieves $<10$\% RMS error when jointly fitting power spectra and mass fractions.

    \item The Mead20+ and Arico24+ models in general achieve a mean RMS error of $\approx$40\% for the inferred profiles (see Fig.~\refc{fig:profile_rms}). The relatively poor accuracy of inferences from the  two models is likely a result of adopted prescriptions and limited flexibility (see Appendix~\refc{appx:fit_all}).
    
\end{itemize}

In the sections that follow we discuss these aspects in more detail.

\subsection{Model ingredients and flexibility}
\label{sec:discussion_flex}

One of the primary goals of this work was to evaluate whether existing halo models are sufficiently flexible to jointly describe power spectra underlying the observable auto- and cross-correlations between probes of matter distribution and thermal properties of gas.

The  percent-level accuracy for the matter--pressure and electron density power spectra (Fig.~\refc{fig:Rk_fit_nene}) is a significant improvement over earlier halo model studies (e.g., \citetalias{Mead2020}), who found it particularly challenging to simultaneously capture the clustering of matter, pressure, and gas density fields. We underscore the importance of slight modifications to the models in making these models more flexible for multi-tracer modeling. For instance, we find it necessary to modify the model presented in \citetalias{Arico2024} by introducing the $T_{\rm w}$ parameter to accurately capture the pressure component.

We also highlight the importance of modeling the one-halo to two-halo transition, and it is worth exploring alternative, more physically motivated approaches for this regime. For example, the Schneider19+ model, which employs untruncated halo profiles accurately recovers the transition-scale power without relying on smoothing i.e., it prefers a smoothing parameter close to unity. These results are also corroborated by other studies which find that using a halo mass definition based on a lower overdensity threshold results in more extended halo profiles and thus reduces the power deficit in the transition regime \citep{Acuto2021}. Incorporating a fully non-linear halo bias prescription might also offer a more natural resolution to this issue \citep{Mead2021b}.

We have demonstrated that achieving percent-level agreement across all three power spectra is feasible with  $\mathcal{O}(15)$ free parameters (see Figs.~\refc{fig:fit_ext},~\refc{fig:fit_mm_mp_mfrac} and~\refc{fig:Rk_fit_nene}). A reduced six-parameter version also performs well when fitting only the matter--matter and matter--pressure spectra (see Fig.~\refc{fig:fit_min}), but the accuracy degrades when including other measurements such as electron density power spectrum or mass fractions. It will be important to identify a minimal set of parameters using a broader range of hydrodynamical simulations that offers sufficient modeling flexibility while minimizing degradation in cosmological constraints. 
These conclusions will also depend on the cosmological probes used and survey-specific details, we defer such investigations to a future study.

\subsection{Physical interpretability of the model}
\label{sec:param_phy}

Beyond achieving accurate fits to data, a key objective of any modeling framework is to remain \textit{physically grounded} rather than merely serving as an \textit{effective} description. 
By this, we mean that the model should retain predictive power for individual contributors to the input summary statistics that we fit to,
 for example, density and pressure profiles across different halo masses or spatial scales. 
While challenging, such models allow us to extract reliable insights from observations and make it easier to isolate and dissect the processes that drive the growth of structure. This emphasis on physical realism is often overlooked, with the primary focus placed on improving accuracy.

In this work, we assess the physical realism of the models by comparing their inferred halo properties to direct measurements from hydrodynamical simulations. We find that models calibrated to power spectra and mass fractions can, in some cases, reasonably recover the scaling relations and thermodynamic profiles as a function of halo mass. However, the accuracy of these inferred properties typically ranges from 10–50\%, depending strongly on the model and the specific data included in the fit. These findings highlight the need for a closer examination of the assumptions and prescriptions underlying these models.

Improving the physical reliability of such models is further complicated by the highly degenerate and multidimensional nature of the parameter space. Nevertheless, several strategies can help mitigate these challenges. We highlight two such approaches below:

\begin{itemize}
    \item Adopting more informative priors to constrain parameters to physically motivated ranges and reducing the potential for compensatory behavior among model components. Constructing and validating such priors requires a suite of large-volume, high-resolution hydrodynamical simulations that span a broad range of subgrid physics implementations.

    \item Requiring model-predicted quantities to satisfy physically motivated constraints. For instance, requiring that temperature increase monotonically with halo mass or that the amount of gas in a halo to be larger than stars. When a model just barely satisfies these constraints it can serve as a diagnostic, much like when a parameter hits the edge of its prior, indicating potential overfitting or a lack of flexibility in the model.  
    Imposing such constraints is also necessary when parameters evolve with halo mass (for e.g., as a power law) for ensuring that the model retains physical meaning for halos at all mass scales.
    
\end{itemize}

We emphasize that the ability of the models to reproduce different thermodynamic profiles of halos in narrow mass bins with reasonable accuracy despite only fitting power spectra is quite remarkable as these summary statistics aggregate contributions across a wide range of masses and spatial scales. 

\subsection{Outlook}

In this work, we have focused on testing these models at the level of the 3D power spectrum to keep the analysis broadly applicable. However, in practice, modeling choices should be guided by the specific characteristics of the surveys they are intended for. Once observational and theoretical systematics as well as the statistical precision of the data are taken into account, some of the limitations highlighted here may prove to be insignificant. On the other hand, numerical artifacts, such a parameter projection effects caused by fitting complex models to under-constraining data, may complicate the interpretation of results compared to the discussion in this paper.

In a follow-up paper, we plan to extend this comparison into observable space by using mock lightcones constructed from simulations. This will allow us to evaluate model performance in a more realistic setting, accounting for observational systematics, projection effects, and the full covariance between different tracers. By working in the data space of specific surveys, we aim to quantify how modeling choices and uncertainties propagate into joint cosmological and astrophysical constraints, ultimately guiding the modeling strategy for upcoming multi-probe datasets.

Multi-wavelength datasets are crucial for mapping out the distribution and thermal properties of cosmic baryons. We have only looked at a handful of probe combinations in this work, but there are many more exciting opportunities  that are worth exploring, such as the low-redshift Lyman-$\alpha$ forest to add another perspective \citep[e.g.,][]{Khaire2024a,Khaire2024b,Tillman2025}.

Cosmological hydrodynamical simulations such as FLAMINGO \citep{Schaye2023} and MillenniumTNG \citep{Pakmor2023} will be instrumental in unlocking the full potential of upcoming multi-wavelength datasets.
Suites of simulations spanning diverse feedback scenarios and cosmologies will enable us to build better priors and rigorously test model performance as well as characterize their accuracy in capturing baryonic feedback as function of background cosmology \citep{RS2025}.

Looking ahead, these analyses will also be key to broadening our understanding of galaxy formation and evolution. While, one approach is to compare the inferred halo properties with results from hydrodynamical simulations, a more fruitful direction is to move beyond empirical scaling relations and develop models of baryonic feedback grounded in first principles. This would involve augmenting the traditional semi-analytic models to co-evolve galaxies and the circumgalactic medium, by directly modeling processes such as energy injection from stellar and AGN feedback, gas cooling, and black hole accretion \citep[e.g.,][]{Pandya2023}. While still in early development, this approach holds strong potential for connecting observations to the physical processes that drives galaxy growth over cosmic time.

\section*{Acknowledgements}
We thank Giovanni Arico for insightful discussions and feedback on the manuscript. We also thank Antonio Ragagnin for assistance with simulation measurements. We also thank Emma Ayçoberry for insightful early discussions that informed the current work. PRS is supported in part by a Joint Fellowship Program between the University of Arizona and CNRS. E.K. and P.R.S. are supported in part by the David and Lucile Packard Foundation. KD acknowledges support by the COMPLEX project from the European Research Council (ERC) under the European Union’s Horizon 2020 research and innovation program grant agreement ERC-2019-AdG 882679. The analyses in this work were carried out using the High Performance Computing (HPC) resources supported by the University of Arizona Technology and Research Initiative Fund (TRIF), University Information Technology Services (UITS), and the office for Research, Innovation, and Impact (RDI) and maintained by the UA Research Technologies Department. The calculations for the hydrodynamical simulations were carried out at the Leibniz Supercomputer Center (LRZ) under the project pr83li (Magneticum).

\section*{Data Availability}

All models are implemented in \textsc{BaryonForge} and are publicly available at \murl{https://github.com/DhayaaAnbajagane/BaryonForge/}. The halo model routines we use are from the CCL library and are available at \murl{https://github.com/LSSTDESC/CCL/}. The data underlying this article will be shared on reasonable request to the corresponding author.



\bibliographystyle{mnras}
\bibliography{example, bib2} 



\appendix

\section{Model Description}
\label{appx:models}
\noindent\textbf{Definitions} All models share a set of basic thermodynamic relations, which we describe below.

The first is the relation between electron pressure $  \mathcal{P}_\text{e}$ and gas pressure $\mathcal{P}_\text{gas}$ assuming fully ionized gas and a Hydrogen mass fraction $X_{\rm H}$
\begin{equation}
\label{eq:Pth_to_Pe}
    \mathcal{P}_\text{e} = \frac{2X_\text{H}+2}{5X_\text{H}+3} \mathcal{P}_\text{gas}.
\end{equation}

This relation can be used to compute the electron number density $n_{\rm e}$ from the gas density $\rho_{\rm gas}$ and the electron temperature $T_{\rm e}$ from from the electron pressure:
\begin{align}
\label{eq:ne}
    n_{\rm e} = \frac{\rho_{\rm gas}}{m_{\rm p}\mu_{\rm e}};\,\,
    T_{\rm e} = \frac{\mathcal{P_{\rm e}}}{n_{\rm e}k_{\rm B}},
\end{align}
where $m_{\rm p}$ is the proton mass, $k_{\rm B}$ is the Boltzmann constant and $\mu_{\rm e}$ is the mean molecular weight per electron 
\begin{equation}
    \mu_{\rm e} = \frac{2}{1+X_{\rm H}}.
\end{equation}
We assume $X_{\rm H}\approx0.75$ and $\mu_{\rm e}\approx1.14$.

\subsection{Mead20+}
\label{appx:M20}
The model assumes that a halo of total mass $M$ is composed of cold dark matter, gas, and stars. In other words the mass fractions of all components sum to unity
\begin{equation}
    f_\text{dm}(M) + f_\text{gas}(M) + f_\text{star}(M) = 1.
\end{equation}

The mass fraction also sets the normalization for the density profile of each individual component i.e.,
\begin{equation}
    \label{eq:HMx_fi}
    f_i(M)M = \int_0^{\infty}4\pi r^2 \rho_i(M,r) dr,
\end{equation}
where $\rho_i(M,r)$  is the density profile for component $i\in\{\text{dm, gas, star} \}$.

The original Mead20 model presented in \citetalias{Mead2020} adopts the virial definition of halos and uses the halo bias and mass function from \citet{Sheth99}, along with the mass–concentration relation from \citet{Duffy2008}. As noted at the beginning of Sec.~\refc{sec:halo_models}, we replace these components with more recent prescriptions from the literature. In addition, we introduce a few other updates to the model, most notably in the treatment of ejected gas. We refer to this updated version as Mead20+ and summarize its details below.

\noindent\textbf{Stars}
The stellar fraction is set following \citet{Fideli2014}
\begin{equation}
    \label{eq:HMx_fstar}
    f_\text{star}(M) = A_*\text{exp}\left[-\frac{\log^2 (M/M_*)}{2\sigma_*^2} \right],
\end{equation}
where $M_*$ determines the characteristic halo mass at which star formation efficiency peaks resulting in a stellar mass fraction $A_*$.
The stellar fraction is further divided into central and satellite components. The stellar fraction in low mass halos ($M<M_*$) is dominated by a central galaxy i.e.,
\begin{equation}
    f_\text{cga}(M) = f_\text{star}(M), \,\,\, f_\text{sga}(M)=0,
\end{equation}
while for high mass halos ($M>M_*$)
\begin{align}
    \label{eq:HMx_fcga}
    f_\text{cga}(M) &= f_\text{star}(M)\left(\frac{M}{M_*}\right)^\eta,\\
    \label{eq:HMx_fsga}
    f_\text{sga}(M) &= f_\text{star}(M) - f_\text{cga}(M),
\end{align}
where $f_\text{cga}$ and $f_\text{sga}$ are the stellar mass fractions of the central and satellite galaxies, respectively. 

In \citetalias{Mead2020} the stellar density profile for central galaxies is modeled as a delta function, here we instead use an exponentially truncated power-law \citep{Schneider2016, Giri2021}

\begin{equation}
\label{eq:rho_cga}
    \rho_\text{cga}(r) = \frac{f_\text{cga}M}{4\pi^{3/2}R_\text{h}}\frac{1}{r^2}\,\text{exp}\left[-\left(\frac{r}{2R_\text{h}^2} \right)^2 \right],
\end{equation}
where $R_\text{h}=0.015\times r_\text{200c}$ sets the cutoff scale. The density profile for satellite galaxies follows an NFW profile
\begin{equation}
    \rho_\text{sga}(r) = \frac{\rho_{\text{sga},0}}{\left(\frac{r}{r_\text{s}}\right)\left(1 + \frac{r}{r_\text{s}}\right)^2},
\end{equation}
where $r_\text{s}$ is the scale radius which is related to the halo radius via the concentration
 $c(M)=r_\mathrm{200c}/r_\mathrm{s}$. The normalization constant $\rho_{\text{sga},0}$ is determined from Eqs.~(\refc{eq:HMx_fi}) and (\refc{eq:HMx_fsga}).

\noindent\textbf{Gas}
Gas is modeled as comprising bound and ejected components. The mass fraction of gas that is retained by halos (i.e. the bound gas) follows the prescription of \citet{Schneider2015}
\begin{equation}
\label{eq:HMx_fBG}
    f_\text{BG}(M) = \frac{\Omega_\text{b}}{\Omega_\text{m}}\frac{1}{1 + (M_0/M)^\beta}.
\end{equation}
The above expression implies that for massive halos ($M\!\gg\!M_0$) the mass fraction of halo gas approaches the cosmic baryon fraction while it transitions to zero in low mass halos ($M\!\ll\!M_0$) with slope $\beta$. The ejected gas fraction is then set by mass conservation\footnote{As noted in \citetalias{Mead2020}, in a scenario where the mass fraction of ejected gas is negative i.e. $f_\text{BG} + f_\text{star}>\Omega_\text{b}/\Omega_\text{m}$, the excess mass is subtracted from the stellar mass fraction.}
\begin{equation}
\label{eq:HMx_fej}
    f_\text{EG}(M) = \frac{\Omega_\text{b}}{\Omega_\text{m}} - f_\text{BG}(M) - f_\text{star}(M).
\end{equation}

The density profile for the bound gas is given by the Komatsu-Seljak (hereafter KS) profile \citep{Komatsu2001,Martizzi2013}
\begin{align}
\label{eq:HMx_BG}
    \rho_{\mathrm{BG}}(r|M) = \rho_{\mathrm{BG},0} \left[\frac{\ln(1+r/r_\mathrm{s})}{r/r_\mathrm{s}}\right]^{1/(\Gamma-1)},
\end{align}
where $\Gamma$ is the polytropic index and $ \rho_{\mathrm{BG},0}$ is the normalization constant.

\citetalias{Mead2020} assume that the ejected gas is is uniformly distributed with the linear matter density field and contributes only to the two halo term. Here, we adopt a more physical physical prescription for the ejected gas profile following \citet{Schneider2015}
\begin{equation}
\label{eq:HMx_rho_EG}
    \rho_\text{EG}(r|M) = \frac{f_{\rm EG}(M)M}{(2\pi R^2_\text{ej})^{3/2}}\, \text{exp}\left[-\frac{1}{2}\left(\frac{r}{R_\text{ej}} \right)^2 \right],
\end{equation}
where $R_\text{ej}$ is obtained by solving 
\begin{equation}
\label{eq:HMx_etab}
    1 - \text{erf}\left[\frac{\eta_\text{b}R_\text{esc}}{\sqrt{2}R_\text{ej}} \right] + \sqrt{\frac{2}{\pi}} \frac{\eta_\text{b}R_\text{esc}}{R_\text{ej}}\text{exp}\left[-\frac{\eta_\text{b}^2R_\text{esc}^2}{2R_\text{ej}^2}\right] = \frac{\Omega_\text{m}}{\Omega_\text{b}}f_\text{EG}(M).
\end{equation}
Here the escape radius is defined as $R_\text{esc}\approx\frac{1}{2}\sqrt{\Delta_\text{200c}}r_\text{200c}$.
The above expression is derived by assuming that the velocity kicks the ejected gas receives from AGN feedback follows Maxwell-Boltzmann statistics and setting the amount of escaped gas (in a timescale $\sim$ Hubble time) equal to the ejected gas fraction \citep[see][for details]{Schneider2015}. The free parameter $\eta_\text{b}$ provides additional flexibility to account for the approximations in calculating $R_\text{esc}$.

\noindent\textbf{Dark Matter}
Motivated by measurements from hydrodynamical simulations \citep[e.g.,][]{Velliscig2014, Mummery2017}, \citetalias{Mead2020} assume that the redistribution of matter caused by baryonic feedback only changes the concentrations while the halo profile shape is still well described by an NFW profile (Eq.~\refc{eq:nfw}). With a halo concentration $c(M)$ measured from a dark-matter only simulation, the halo concentration in the presence of baryonic feedback is parameterized as
\begin{align}
    \label{eq:HMx_cmod}
    c(M) \rightarrow c(M) \left[1 + \epsilon_1 +(\epsilon_2-\epsilon_1)\frac{f_{\mathrm{bnd}(M)}}{\Omega_\mathrm{b}/\Omega_\mathrm{m}} \right],
\end{align}
here parameters $\epsilon_1$ and  $\epsilon_2$ govern the modification in halo concentration for low- and high-mass halos, respectively, and act as a link between the baryonic components and dark matter. Note that the modified concentration is also used when computing the gas density and temperature profiles.

\noindent\textbf{Pressure}
The Mead20+ model assumes that the total gas pressure and density are related to each other with a polytropic equation of state i.e., $\mathcal{P}\propto\rho^\Gamma$. The resulting gas density profile is given by Eq.~(\refc{eq:HMx_BG}) and the gas temperature profile is expressed as \citep{Komatsu2001}
\begin{equation}
    T(r|M) = T_{\mathrm{200c}} \frac{\ln(1+r/r_\mathrm{s})}{r/r_\mathrm{s}}.
\end{equation}
The halo temperature is defined as
\begin{equation}
\label{eq:HMx_Tv}
        T_{\mathrm{200c}} = \alpha \frac{G M m_{\rm p} \mu_{\rm p}}{\frac{3}{2}k_Br_{\mathrm{200c}}} (1+z),
\end{equation}
where $\alpha$ is a free parameter that accounts for non-thermal pressure support. Note that this amplitude rescaling of the pressure profile is too simplistic since the non-thermal contribution has a different radial dependence and also varies halo mass \citep{Nelson2014}.

The pressure contribution from the ejected gas is computed assuming that it is heated to a constant temperature of $T_\text{w}$, the temperature of the warm-hot intergalactic medium, which we treat as a free parameter.

The total electron pressure is computed as a sum of contributions from the bound and ejected gas components via the ideal gas law and assuming complete ionization
\begin{align}
\label{eq:HMx_Pe}
    \mathcal{P}_\text{e}(r|M) = \frac{\rho_{\text{BG}}(r)}{m_\text{p} \mu_\text{e}}k_\text{B} T(r) + \frac{\rho_{\text{EG}}(r)}{m_\text{p} \mu_\text{e}}k_\text{B} T_\text{w}.
\end{align}

\subsection{Schneider19+}
\label{appx:S19}

In this formalism, the total matter density ($\rho_\text{m}$) is expressed as a sum of three components: the central galaxy ($\rho_\text{cga}$), gas ($\rho_\text{gas}$), and collisionless matter ($\rho_\text{clm}$)
\begin{equation}
\label{eq:god_rhom}
    \rho_\text{m}(r) = \rho_\text{cga}(r) + \rho_\text{gas}(r) + \rho_\text{clm}(r),
\end{equation}
where $\rho_\text{clm}$ includes contributions from both dark matter and  satellite galaxies. For a component $i\in\{\text{cga,\,gas,\,clm}\}$, the mass enclosed within a sphere of radius $r$ is given by
\begin{equation}
    M_i(<r)=\int_0^r 4\pi y^2 \rho_i(y) \text{d}y.
\end{equation} 
The total mass contained in a halo is then expressed as
\begin{equation}
    M_\text{tot} = M_\text{cga}(<\infty) + M_\text{gas}(<\infty)  + M_\text{clm}(<\infty).
\end{equation}

We now summarize the modeling assumptions and ingredients.

\noindent\textbf{Star}
The stellar component comprises contribution from the central and satellite galaxies. The
total stellar fraction and central galaxy fraction are parameterized as
\begin{align}
\label{eq:god_fstar}
f_\text{star}(M) &= \frac{2A}{\left(\frac{M}{M_1} \right)^\tau + \left(\frac{M}{M_1} \right)^\eta},\\
\label{eq:god_fcga}
f_\text{cga}(M) &= \frac{2A}{\left(\frac{M}{M_1} \right)^{\tau+\tau_\delta} + \left(\frac{M}{M_1} \right)^{\eta+\eta_\delta}},
\end{align}
where $\tau, \tau_\delta, \eta, \eta_\delta, \text{and}\, M_1$ are free parameters. . Note that the double power law is necessary to maintain realistic stellar fractions in low mass halos \citep{Moster2013} and is different from what was originally presented in \citet{Schneider2019, Pandey2024}. The abundance of satellite galaxies is obtained as $f_\text{sga} =f_\text{star}-f_\text{cga}$.

The stellar density profile of the central galaxy is modeled as a power-law with an exponential cutoff as expressed in Eq.~(\refc{eq:rho_cga}).

\noindent\textbf{Gas}
The gas is modeled as a single bound component with mass fraction set by
\begin{equation}
    \label{eq:god_fgas}
    f_\text{gas}(M) = \frac{\Omega_\text{b}}{\Omega_\text{m}} - f_\text{star}(M).
\end{equation}
The gas distribution is given by an untruncated density profile following the prescription of \citet{Giri2021}
\begin{equation}
    \label{eq:god_rhogas}
   \rho_\text{gas}(r|M) = \frac{\rho_\text{gas,0}}{\left[1 + \left(\frac{r}{\theta_\text{co} r_\text{200c}}\right)^\beta\right] \left[1 + \left(\frac{r}{\theta_\text{ej} r_\text{200c}}\right)^\gamma\right]^{\frac{\delta - \beta}{\gamma}}},
\end{equation}
where $\{\beta,\delta,\gamma\}$ control the slope of the profile, $\{\theta_\text{co},\theta_\text{ej}\}$ define the core and ejection radii, respectively, and $\rho_\text{gas,0}$ is the normalization constant. The parameter $\beta$ depends on halo mass as
\begin{equation}
\label{eq:god_beta}
\beta = \frac{3 \left(M / M_c \right)^{\mu_\beta}}{1 + \left(M / M_\text{c} \right)^{\mu_\beta}},
\end{equation}
where $M_\text{c}$ is the halo mass below which the gas profile becomes
shallower than the NFW profile and $\mu_\beta$ determines the mass scaling of the slope. The parameter $M_\text{c}$ is also evolved with redshift as
\begin{equation}
    M_\text{c} = M_\text{c,0} (1+z)^{\nu_{M_\text{c}}},
\end{equation}
with free parameters $M_\text{c,0}$ and $\nu_{M_\text{c}}$. The remaining slope parameters $\delta$ and $\gamma$ do not have an explicit mass or redshift dependence and only their amplitude is varied to reduce degeneracies.

In \citet{Pandey2024} he parameters $\theta_\text{co}$ and $\theta_\text{ej}$ are also expressed in terms of halo mass, halo concentration, and redshift
\begin{equation}
\label{eq:god_theta}
    \theta_\text{co/ej} = \theta_\text{co/ej,0}\left(\frac{M}{M_\text{co/ej}}\right)^{\mu_\text{co/ej}} \left(\frac{1}{c_\text{200c}} \right)^{\zeta_\text{co/ej}} (1+z)^{\nu_\text{co/ej}},
\end{equation}
where $M_\text{co/ej}$ determine the scaling with mass, we fix both parameters to $10^{16}\,h^{-1}{\rm M}_\odot$.  We do not consider variations with halo concentration or redshift and fix $\zeta_\text{co/ej}$ and $\nu_\text{co/ej}$ to zero.

\noindent\textbf{Collisionless matter}
\label{sec:god_clm}
The collisionless matter comprises the dark matter and satellite galaxies components. The mass fraction of collision matter is therefore
\begin{equation}
    f_\text{clm}(M)= 1 - \frac{\Omega_\text{b}}{\Omega_\text{m}} + f_\text{sga}(M).
\end{equation}

In the absence of baryonic effects, dark matter is described by a truncated NFW profile (Eq.~\refc{eq:tnfw}).

The baryonic components of a halo are redistributed by astrophysical processes and thereby result in the contraction or expansion of collisionless matter via gravitational effects. The modification is modeled by displacing a sphere of initial radius $r_\text{i}$ and mass $M_\text{i}$ to a final radius $r_\text{f}$ and mass $M_\text{f}$ \citep{Schneider2016,Schneider2019}. Motivated by numerical simulations \citep{Abdadi2010}, these quantities are related as
\begin{equation}
\label{eq:god_relax}
    \psi=\frac{r_\text{f}}{r_\text{i}} = 1 + a_\psi\left[\left(\frac{M_\text{i}}{M_\text{f}} \right)^{n_\psi} - 1 \right],
\end{equation}  
where $a_\psi=0.3$ and $n_\psi=2$. The mass contained within the initial and final shells is given by
\begin{align}
\label{eq:god_Mi}
    M_\text{i} &= M_\text{tnfw}(<r_\text{i}),\\
    \label{eq:god_Mf}
    M_\text{f} &= f_\text{clm}M_\text{tnfw}(<r_\text{i}) + M_\text{cga}(<r_\text{f}) + M_\text{gas}(<r_\text{f}).
\end{align}
The density profile of the relaxed collisionless matter $\rho_\text{clm}(r)$ is obtained by iteratively solving Eq.~(\refc{eq:god_relax}) for $\psi$ and is expressed as
\begin{equation}
     \rho_\text{clm}(r) = \frac{f_\text{clm}(M)}{4\pi r^2} \frac{{\rm d}}{{\rm d}r}M_\text{tnfw}(r/\psi).
\end{equation}

\noindent\textbf{Pressure}
The total pressure in a halo is obtained by assuming hydrostatic equilibrium 
\begin{equation}
    \frac{1}{\rho_\text{gas}}\frac{d\mathcal{P}_\text{tot}}{dr} = -\frac{GM(<r)}{r^2},
\end{equation}
where $M(<r)=\int_0^r4\pi y^2\rho_\text{m}\text{d}y$ and $\rho_\text{m}$ is the total matter density given by Eq.~(\refc{eq:god_rhom}).
The above expression includes contributions from both thermal and non-thermal pressure, the latter is parameterized as \citep{Shaw2010,Osato2023}
\begin{equation}
\label{eq:god_Rnt}
    R_\text{nt} = \frac{\mathcal{P}_\text{nt}}{\mathcal{P}_\text{tot}} = \alpha_\text{nt}f(z)\left(\frac{r}{r_\text{200c}}\right)^{\gamma_\text{nt}},
\end{equation}
where $\alpha_\text{nt}$ determines the amplitude of non-thermal pressure support and $\gamma_\text{nt}$ is the slope of the radial profile. The function $f(z)$ governing the redshift evolution is given by
\begin{equation}
    f(z) = \text{min}[(1+z)^{\nu_\text{nt}}, (f_\text{max}-1)\text{tanh}(\nu_\text{nt}z) +1],
\end{equation}
where $f_\text{max}=6^{-\gamma_\text{nt}}/\alpha_\text{nt}$ so that $R_\text{nt}<1$ in the radial range $0<r<6r_\text{200c}$ \citep{Shaw2010}. The thermal pressure $\mathcal{P}_\text{th}$ is then
\begin{equation}
    \mathcal{P}_\text{th} = \mathcal{P}_\text{tot}\times[0, 1-R_\text{nt}].
\end{equation}

The electron pressure, which sources the tSZ signal, is computed from $\mathcal{P}_\text{th}$ following Eq.~(\refc{eq:Pth_to_Pe}). The electron temperature profile is obtained via the ideal gas law as expressed in Eq.~(\refc{eq:ne}).


\subsection{Arico24}
\label{appx:A20}

In this model, which we refer to as Arico24, the halo comprises three main components: the central galaxy, gas, and collisionless matter --- where the latter includes both dark matter and the stellar content of satellite galaxies.

Below we provide a summary for completeness but we refer the reader to the aforementioned papers for details.

\noindent\textbf{Star}
The mass fraction of the central galaxy is modeled following \citet{Behroozi2013}
\begin{equation}
\label{eq:AA_fcga}
    f_\text{cga}(M_{200}) = \epsilon \left(\frac{M_1}{M_{200}}\right) 10^{g(\log_{10}(M_1/M_{200})) - g(0)},
\end{equation}
where $g(x)$ is a fitting function expressed as
\begin{equation}
\label{eq:AA_gx}
    g(x) = -\log(10^{\alpha x}+1) + \delta \frac{\log (1+e^x)^\gamma}{1 + {\rm exp}(10^{-x})}.
\end{equation}

The parameter $M_1$ sets the characteristic halo mass for which the mass fraction of the central galaxy is $\epsilon$ and its value at $z=0$, $M_{1,0}$, is a free parameter. The remaining parameters in the expression $\{\alpha,\,\delta,\,\gamma\}$ are linearly dependent on the redshift and their evolution is fixed to the best-fit values reported in \citet{Behroozi2013} and \citet{Kravtsov2018} (also summarized in Appendix A of \citealt{Arico2020}).

The density profile for the central galaxy is modeled as an exponentially truncated power-law
\begin{equation}
\label{eq:AA_rhocga}
    \rho_\text{cga}(r) = \frac{\rho_{\text{cga},0}}{R_\text{h}r^{\alpha_\text{g}}}\, \text{exp}\left[ -\left(\frac{r}{2R_\text{h}} \right)^2 \right],
\end{equation}
where $\alpha_g$ is the inner slope of the density profile (fixed to 2), $R_\text{h}$ is the stellar half-light radius and is fixed to $0.015\,\times\, r_\text{200c}$, and
$\rho_{\text{cga},0}$ is a normalization constant that ensures the volume integrated profile gives the correct mass fraction.

The mass fraction of the satellite galaxies $f_\text{sga}$ is also obtained via Eqs.~(\refc{eq:AA_fcga}) \& (\refc{eq:AA_gx}) but after linearly scaling the best-fit values by a free parameter \citep{Watson2013}, e.g., $M_{1,0}\rightarrow M_\text{1,sat} M_{1,0},\,\epsilon\rightarrow\epsilon_{\rm sat}\epsilon,\, \alpha\rightarrow\alpha_{\rm sat}\alpha$, etc. The density profile of satellite galaxies, which are effectively collisionless, is not modeled separately. Instead, their contribution is incorporated into the dark matter profile.

\noindent\textbf{Gas}
The total gas fraction $f_\text{gas}$ is the fraction of baryons that have not been converted into stars
\begin{equation}
\label{eq:AA_fgas}
    f_\text{gas}(M) = \frac{\Omega_\text{b}}{\Omega_\text{m}} - f_\text{cga}(M)- f_\text{sga}(M).
\end{equation}
The total gas fraction is divided into three components: bound gas, reaccreted gas, and gas ejected beyond the halo. The first two together make up what is referred to as the halo gas. The mass fractions of the halo ($f_\text{HG}$) and reaccreted ($f_\text{RG}$) components are defined as follows
\begin{align}
\label{eq:AA_fHG}
    f_\text{HG}(M) &= f_\text{BG}(M) + f_\text{RG}(M) = \frac{f_\text{gas}}{1 + (M_\text{c}/M_{200})^\beta},\\
    \label{eq:AA_fRG}
    f_\text{RG}(M) &= \frac{f_\text{gas}-f_\text{HG}}{1 + (M_r/M_{200})^{\beta_r}} = f_\text{HG}\frac{(M_\text{c}/M)^\beta}{1 + (M_r/M)^{\beta_r}}.
\end{align}

Mass conservation then sets the fraction of the bound and ejected components
\begin{align}
    f_\text{BG}(M) &= f_\text{HG}(M) - f_\text{RG}(M),\\
    f_\text{EG}(M) &= f_\text{gas}(M) - f_\text{HG}(M).
\end{align}

The modeling of the reaccreted gas component was initially motivated by the findings of \citet{Foreman2020}, however  \citet{Arico2021} do not find any preference for this component based on fits to hydrodynamical simulations. Therefore, we fix $M_\text{r}=10^{18}M_\odot\,h^{-1}$  which effectively sets the reaccreted gas mass fraction to zero for the relevant halo masses according to Eq.~(\refc{eq:AA_fRG}).

The density profile for each component is described below

    \begin{itemize}
        \item Bound gas: The density profile of the bound gas is represented as
    \begin{equation}
    \label{eq:AA_rhoBG}
        \rho_\text{BG}(r\leq r_\text{out}) =  \frac{\rho_{\text{BG},0}}{\left(1 + \frac{r}{\theta_\text{inn}r_\text{200c}}\right)^{\beta_\text{i}}\left(1 + \left( \frac{r}{\theta_\text{out}r_\text{200c}}\right)^2\right)^2},
    \end{equation}
    where $\rho_0$ is a normalization constant and $\theta_\text{inn/out}$ set the radii at which the inner/outer slope changes. Beyond the outer radius $r_\text{out}=\theta_\text{out}\times r_\text{200c}$ the gas perfectly traces an NFW profile. The inner slope $\beta_\text{i}$ is given by
    \begin{equation}
    \label{eq:AA_betai}
        \beta_\text{i} = 3 - \left(\frac{M_\text{inn}}{M}\right)^{\mu_\text{i}}.
    \end{equation}
    Here, $M_\text{inn}$ is the characteristic halo mass at which the inner slope matches the NFW profile and $\mu_\text{i}$ captures its variation with halo mass.

    \item Ejected gas: The gas ejected from halos by AGN feedback is modeled following Eq.~(\refc{eq:HMx_rho_EG}).
    The gas ejection radius $R_\text{ej}$ is expressed in terms of the free parameter $\eta$ as $R_\text{ej}=\eta\,\times\,0.75R_\text{esc}$, where $R_\text{esc}\approx\frac{1}{2}\sqrt{\Delta_\text{200c}}r_\text{200c}$ is the  halo escape radius.
    \item Reaccreted gas: The reaccreted gas component is parameterized as
    \begin{equation}
    \label{eq:AA_rho_RG}
        \rho_\text{RG}(r) =   \frac{\rho_{\text{RG},0}}{\sqrt{2 \pi} \sigma_\text{r}} \exp\left[-\frac{(r - \mu_\text{r})^2}{2 \sigma_\text{r}^2}\right],
    \end{equation}
    where $\mu_\text{r}=0.3\,\times\,r_\text{200c}$ and $\sigma_\text{r}=0.1\,\times\,r_\text{200c}$ set the location and extent of the spatial distribution, and $\rho_0$ is a normalization constant.
    \end{itemize}

\noindent\textbf{Dark matter}
\label{sec:AA_DM}
The distribution of the dark matter component, including the contribution from satellites, is modeled as a piece-wise function
\begin{equation}
\label{eq:AA_pnfw}
    \rho_\text{dm}(r) =     \begin{cases}
     \frac{\rho_{\text{dm},0}}{(r/r_\text{s})(1+r/r_\text{s})^2} & r\leq r'\\
       \rho_\text{dm}(r_{200}) & r'<r\leq r_{200} \\
      0 & r>r_{200}
    \end{cases}
\end{equation}
The density profile is forced to match at $r'$ and $r_\text{200c}$ to ensure continuity (refer to Appendix A in \citealt{Arico2020} for details).

The redistribution of baryons has a gravitational back-reaction on the dark matter particles, which otherwise approximately trace an NFW density profile. This halo relaxation is modeled using the same approach as for the Schneider19+ model (Sec.~\refc{sec:god_clm}) and Eqs.~(\refc{eq:god_Mi}) \& (\refc{eq:god_Mf}) are evaluated using the profile ingredients described above.

\noindent\textbf{Pressure}
The total pressure is computed assuming that the gas follows a polytropic equation of state \citep{Komatsu2001}
\begin{equation}
\label{eq:AA_PBG}
    \mathcal{P}_\text{BG}(r) = \mathcal{P}_0 \cdot \rho_\text{BG}(r)^\Gamma,
\end{equation}
where the normalization constant $\mathcal{P}_0$ is given by
\begin{equation}
        \mathcal{P}_0 = \frac{4\pi G\rho_\text{DM}(r_\text{s})r_\text{s}^2}{\rho_{\text{BG},0}^{\Gamma-1}}\frac{\Gamma-1}{\Gamma}.
\end{equation}
In the above expression $G$ is the gravitational constant and $r_\text{s}$ is the halo scale radius. The polytropic index $\Gamma$ is modeled as a function of the halo concentration $c$ and $\theta_\text{out}$
\begin{align}
    \Gamma = 1 + \frac{(1 + x')\ln(1+x')-x'} {(1+3x')\ln(1+x')}; \, \, x'=c_{\rm 200c}\times\theta_\text{out}.
\end{align}

The contribution of non-thermal pressure support to the total pressure in Eq.~(\refc{eq:AA_PBG}) is modeled as a function of the peak height using the fitting function of \citet{Green2020}. The thermal pressure is written as
\begin{equation}
    \mathcal{P}_\text{BG, th}(r) = f_\text{th}(r/r_\text{200m})\mathcal{P}_\text{BG}(r).
\end{equation}
The fraction of thermal pressure support is
\begin{equation}
        f_\text{th}(x) = A \left(1+e^{-(x/B)^C} \right)(\nu_\text{200m}/4.1)^{D/(1+(x/E)^F)},
\end{equation}
where, $x=r/r_\text{200m}$ and $\nu_\text{200m}$ is the peak height. The parameters $\{B, C, D, E,F\}$ are fixed to their best-fit values presented in \citet{Green2020}. For additional flexibility, $A$ is further expressed as function of two free parameters, the amplitude $A_{\rm th}$ and redshift dependence $\alpha_\text{th}$,
\begin{equation}
\label{eq:AA_Ant}
    A = A_{\rm th}(1+z)^{\alpha_\text{th}}.
\end{equation}

The electron temperature profile is computed by imposing the ideal gas law as expressed in Eq.~(\refc{eq:ne}).

Note that in \citetalias{Arico2024}, the ejected gas follows the  temperature profile of the bound gas. In our preliminary analysis we found that this choice made it difficult for the model to match the amplitude of the matter--pressure power spectrum on large scales. We therefore set the temperature of the ejected gas independently using the $T_\text{w}$ parameter\footnote{None that the $T_\text{w}$ parameter in our implementation is distinct from the  $T_\text{field}$ parameter in \citetalias{Arico2024}, where the latter sets the temperature of the \textit{field} component i.e., gas particles in an N-body simulation that are not associated with any halo.} (analogous to the $T_{\rm w}$ parameter in the Mead20+ model).

\section{Measurements from simulations}
\subsection{Power spectrum}
\label{appx:Pk_magneticum}
We use the Python library \textsc{Pylians} \citep{Pylians} to estimate the power spectrum from the particle data. We discretize the field onto a 3D grid with 1024$^3$ voxels using cloud-in-cell (CIC) mass assignment scheme. The default $P(k)$ measurements are in linearly spaced $k$-bins, with bin size equal to the fundamental model i.e., $\Delta k=2\pi/L=2\pi/(640\,h^{-1}{\rm Mpc}) \approx 0.01\,{\rm Mpc}^{-1}h$. To suppress the noise on large-scales, we re-bin all the measured power spectra in 100 uniformly linearly-spaced bins from $k\sim0.03-4{\rm Mpc}^{-1}h$. All our power spectrum measurements are in comoving units.

When estimating the electron pressure power spectrum from the simulation we do not select gas particles that are at very low temperatures and are thus likely to comprise the cold neutral component. Specifically, we exclude gas particles with temperature $<3\times10^4$ K and cold gas fraction $>0.1$ \citep{Borgani2004}. Although, we find that this choice has no impact on our measurements.

\subsection{Halo profiles and mass fractions}
\label{appx:Prof_magneticum}

To measure halo profiles, we first divide the halo population into four uniformly log-spaced mass bins spanning the range 
$M_{\rm 200c}=10^{13}-10^{15}\,h^{-1}{\rm M}_\odot$. We then use the halo catalog to randomly select 200 objects in each mass bin.

Radial profiles are measured in 16 logarithmically spaced bins from $R=0.08-3\,R_{\rm 200c}$. For the temperature profile, we compute the mass-weighted mean temperature of gas particles within each radial shell.

We also measure the bound gas and stellar mass fractions within 
$R_{\rm 200c}$, defined as:
\begin{align}
f_{\rm gas} &= \frac{M_{\rm gas}(<R_{\rm 200c})}{M_{\rm tot}(<R_{\rm 200c})}, \\
f_{\rm star} &= \frac{M_{\rm star}(<R_{\rm 200c})}{M_{\rm tot}(<R_{\rm 200c})},
\end{align}
where $M_i(<R_{\rm 200c})$ denotes the total mass of particles of type 
$i$ (gas or stars) within a sphere of radius 
$R_{\rm 200c}$
 , and 
$M_{\rm tot}(<R_{\rm 200c})$ includes all particle types: gas, stars, black holes, and dark matter.

We compute the mean profiles (and mass fractions) in each mass bin as the average of all halos in the mass bin. When making model predictions for these quantities, we evaluate them at the same halo masses as the measurement and similarly obtain the average in a mass bin.

\section{Constraints on stellar fractions}
\label{appx:fstar_constraint}
\begin{figure*}
    \centering
    \includegraphics[width=0.3\linewidth]{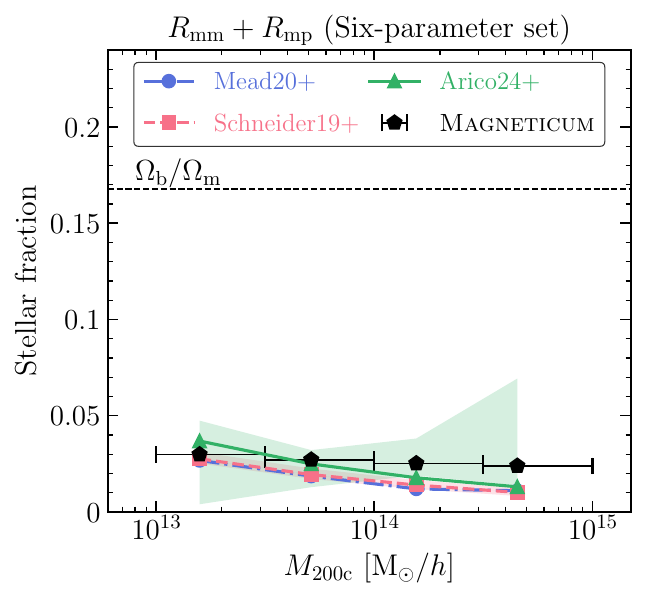}
    \includegraphics[width=0.3\linewidth]{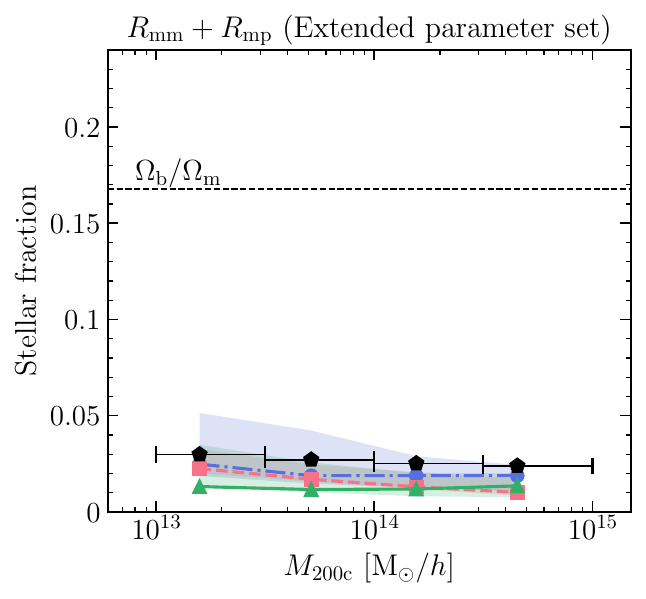}
    \includegraphics[width=0.3\linewidth]{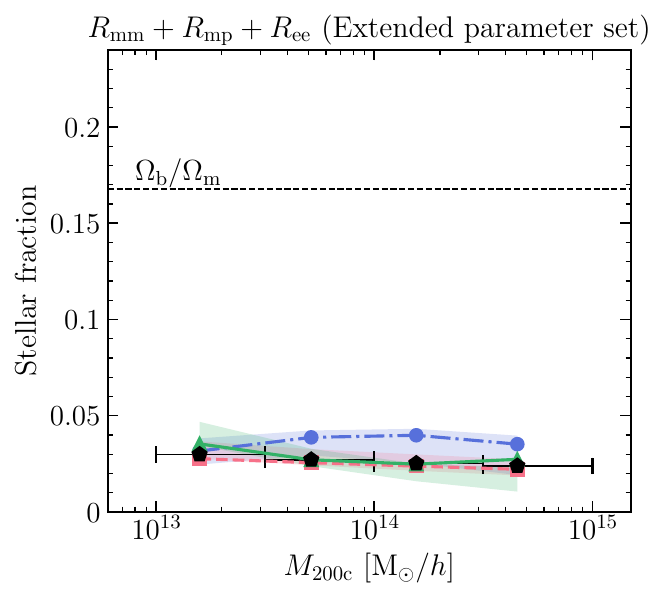}
    \caption{Predicted stellar abundance as a function of halo mass. Panels from left to right show prediction from the models fit to $R_{\rm mm} + R_{\rm mp}$ with the six parameter set (Sec.~\protect\refc{sec:res_min}), $R_{\rm mm} + R_{\rm mp}$ with the extended set (Sec.~\protect\refc{sec:mm_mp_ext}), and $R_{\rm mm} + R_{\rm mp} +  R_{\rm ee}$ with the extended parameter set (Sec.~\protect\refc{sec:mm_mp_nene}), respectively.}
    \label{fig:result_fstar}
\end{figure*}

In this appendix we present the inferred constraints on the stellar fractions, the results are presented in Figure~\refc{fig:result_fstar}.
We find that all models underpredict the stellar content of halos, particularly at the high mass end, when only fitting to $R_{\rm mm}$ and $R_{\rm mp}$ (left and middle panel). Including information about the gas distribution through $R_{\rm ee}$ significantly improves the inferred stellar fractions, bringing them into better agreement with simulation measurements (right panel).

\section{Fits to power spectra, mass fractions and thermodynamic profiles}
\label{appx:fit_all}
\begin{figure*}
    \centering
    \includegraphics[width=\linewidth]{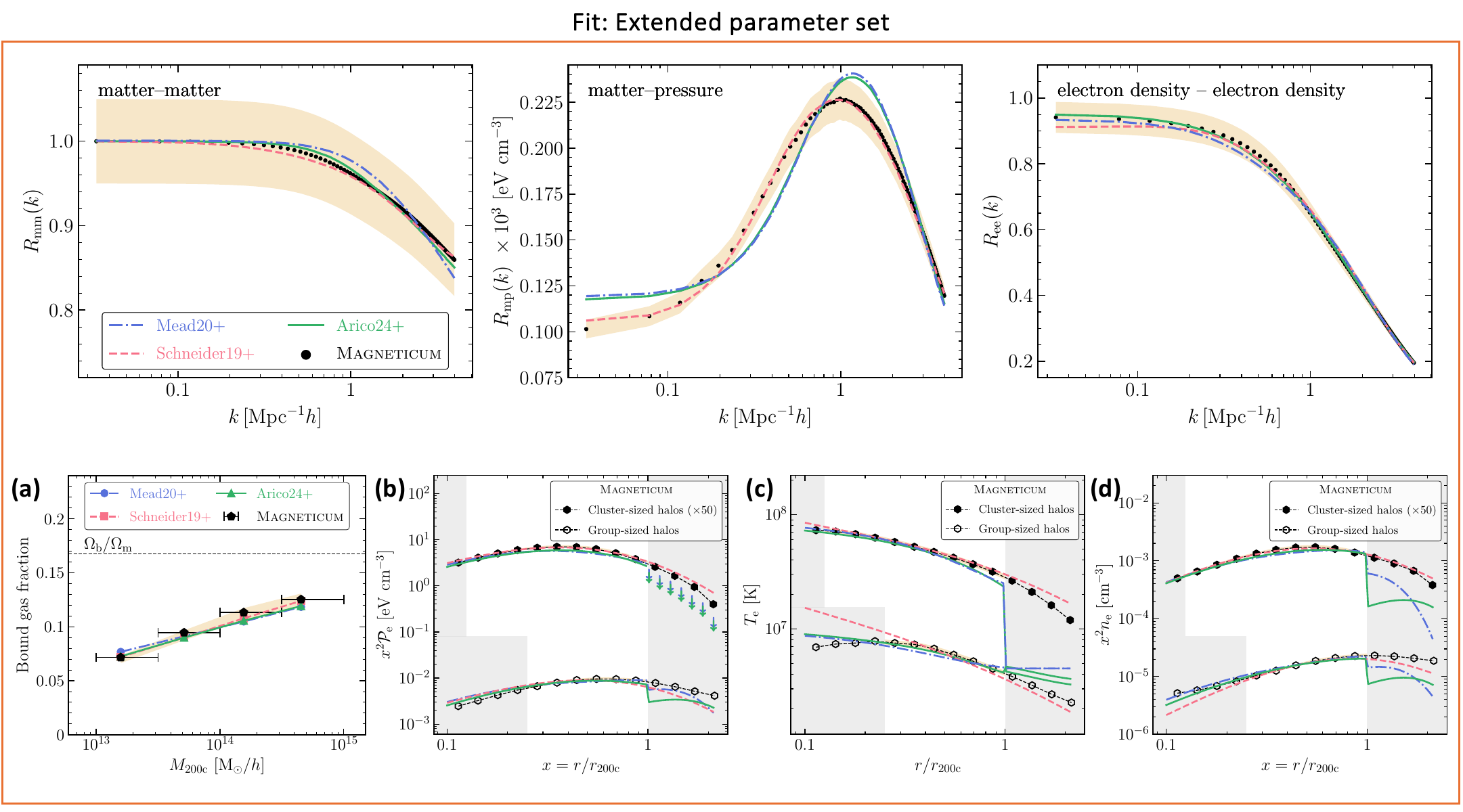}
    \caption{Same as Fig. \protect\refc{fig:fit_min} but fitting $R_{\rm mm}(k)$, $R_{\rm mp}(k)$, $R_{\rm ee}(k)$, $f_{\rm BG}$, $f_{\rm star}$, $\mathcal{P}_{\rm e}(r)$, $T_{\rm e}(r)$, $\rho_{\rm e}(r)$ with an extended parameter. The gray shaded regions in panels (b) through (d) indicate the scales not included in the fit. Note that the fits for the stellar mass fraction $f_{\rm star}$ are not shown.}
    \label{fig:res_mm_mp_nene_mfrac_prof}
\end{figure*}
In Sec.~\refc{sec:results}, we found that while all models accurately reproduce the power spectra and mass fractions, they fail to predict the thermodynamic profiles with similar precision. In this appendix, we test whether the models possess sufficient flexibility to converge to the correct solution when all the data are fit jointly. 

Specifically, we fit (a) the power spectra of matter, pressure, and electron density; (b) the bound gas and stellar mass fractions, and (c) the radial profiles of electron pressure ($\mathcal{P}_{\rm e}$), temperature ($T_{\rm e}$), and density ($\rho_{\rm e}$). The fits are performed using the extended parameter set comprising $\mathcal{O}(15)$ parameters (see Table~\refc{tab:model_params}). The likelihood contribution of the and mass fractions and power spectra is detailed in Sec.~\refc{sec:analysis_setup} (Eqs.~\refc{eq:like_mfrac} and~\refc{eq:like_response}, respectively).

The likelihood contribution of the profiles is
\begin{equation}
    \log\mathcal{L}_{\rm profiles} = -\frac{1}{2}\left[\sum_i\left(\frac{\mathcal{X}_i^{\rm sim}(r)-\mathcal{X}_i^{\rm model}(r)}{\sigma_i(r)}\right)^2_{\rm group}  +\sum_i\left(\frac{\mathcal{X}_i^{\rm sim}(r)-\mathcal{X}_i^{\rm model}(r)}{\sigma_i(r)}\right)^2_{\rm cluster}\right],
\end{equation}
where, $i=\{\mathcal{P}_{\rm e}, T_{\rm e}, \rho_{\rm e}\}$ and the superscript denotes if the profile is measured from the simulation or a model prediction. The two terms on the right hand side correspond to the profiles of  group-mass ($13\leq\log (M_{\rm200c}h/\mathrm{M}_\odot) <13.5$) and cluster-mass ($14.5\leq\log (M_{\rm200c}h/\mathrm{M}_\odot)<15$) halos, respectively. We fit the profiles over the radial range $x_{\rm min} < r/r_{\rm 200c} \lesssim 1$, with $x_{\rm min} = 0.25$ for group-mass halos and 0.12 for cluster-mass halos to maintain consistency in the sensitivity to physical scales probed in previous fits (Figs. \refc{fig:fit_min}, \refc{fig:fit_ext}, \refc{fig:fit_mm_mp_mfrac}, and \refc{fig:Rk_fit_nene}). For all profiles we assume a measurement uncertainty $\sigma_i(r)=5$\%.

The results are presented in Fig.~\refc{fig:res_mm_mp_nene_mfrac_prof}. We find that the Schneider19+ model provides a reasonable match to all profiles across both mass bins. The power spectra and gas mass fractions are also fit with comparable accuracy to the previous cases where the profiles were not included in the fit. The Mead20+ and Arico24+ models similarly reproduce most quantities well, with the exception of the matter--pressure response, which is matched only at the 5--10\% level. This suggests that these models are not flexible enough to simultaneously describe the halo profiles and power spectra. Note that all models fit the stellar mass fractions to within a few percent; these are therefore not shown.

\section{Parameter posteriors}
\label{appx:param_posteriors}

\begin{figure*}
    \centering
    \includegraphics[width=\linewidth]{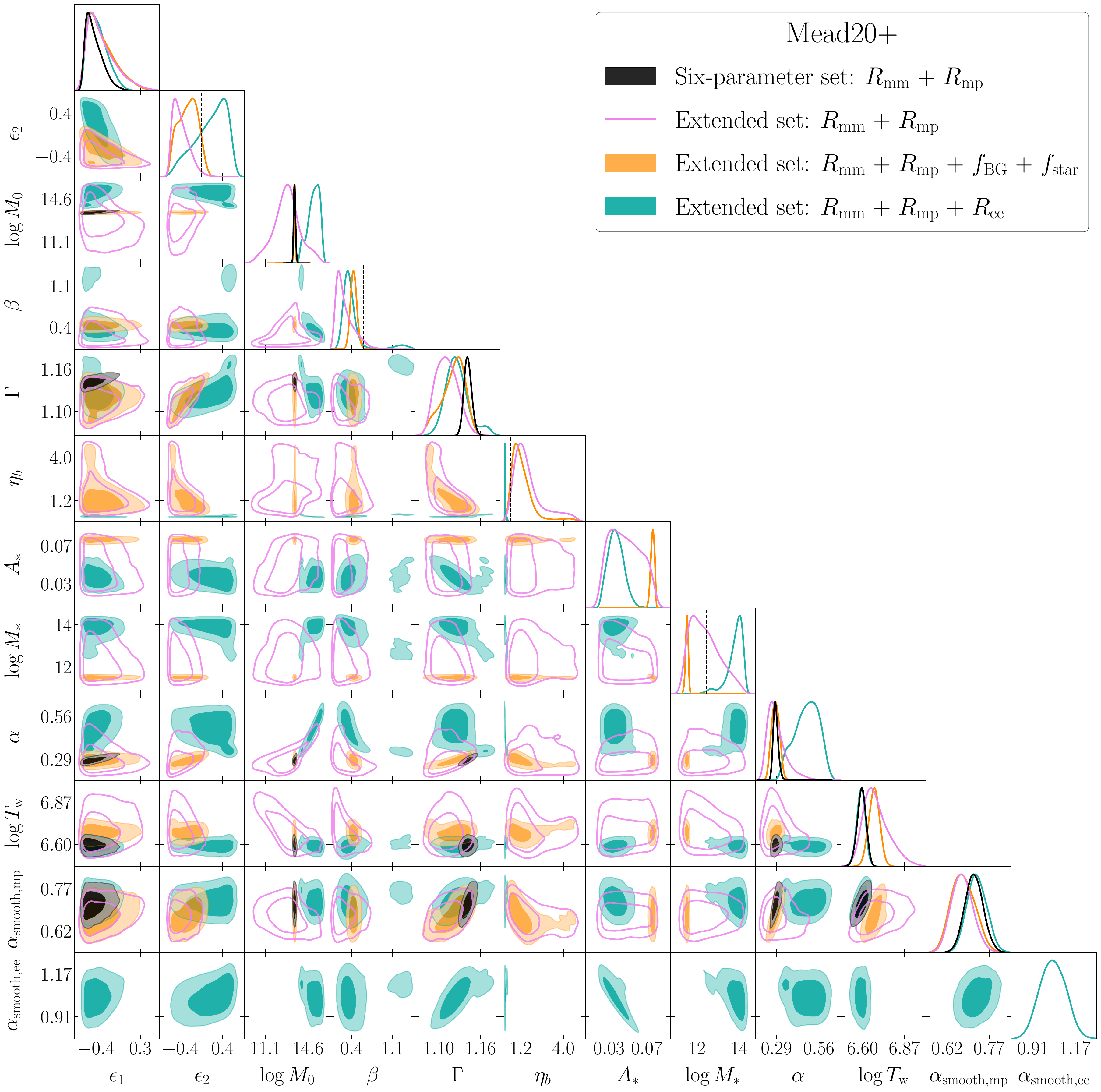}
    \caption{Parameter posteriors for the Mead20+ model when fitting to the \textit{Magneticum} simulation at $z=0.25$. We show marginalized contours for different combinations of the matter--matter, matter--pressure, and electron density power spectrum response along with the bound gas and stellar fractions. The best-fit parameters for each case are listed in Table \protect\refc{tab:bf_params}. The dashed black line indicates the fiducial values of the parameters that are held fixed in the six-parameter set.}
    \label{fig:post_M20}
\end{figure*}

\begin{figure*}
    \centering
    \includegraphics[width=\linewidth]{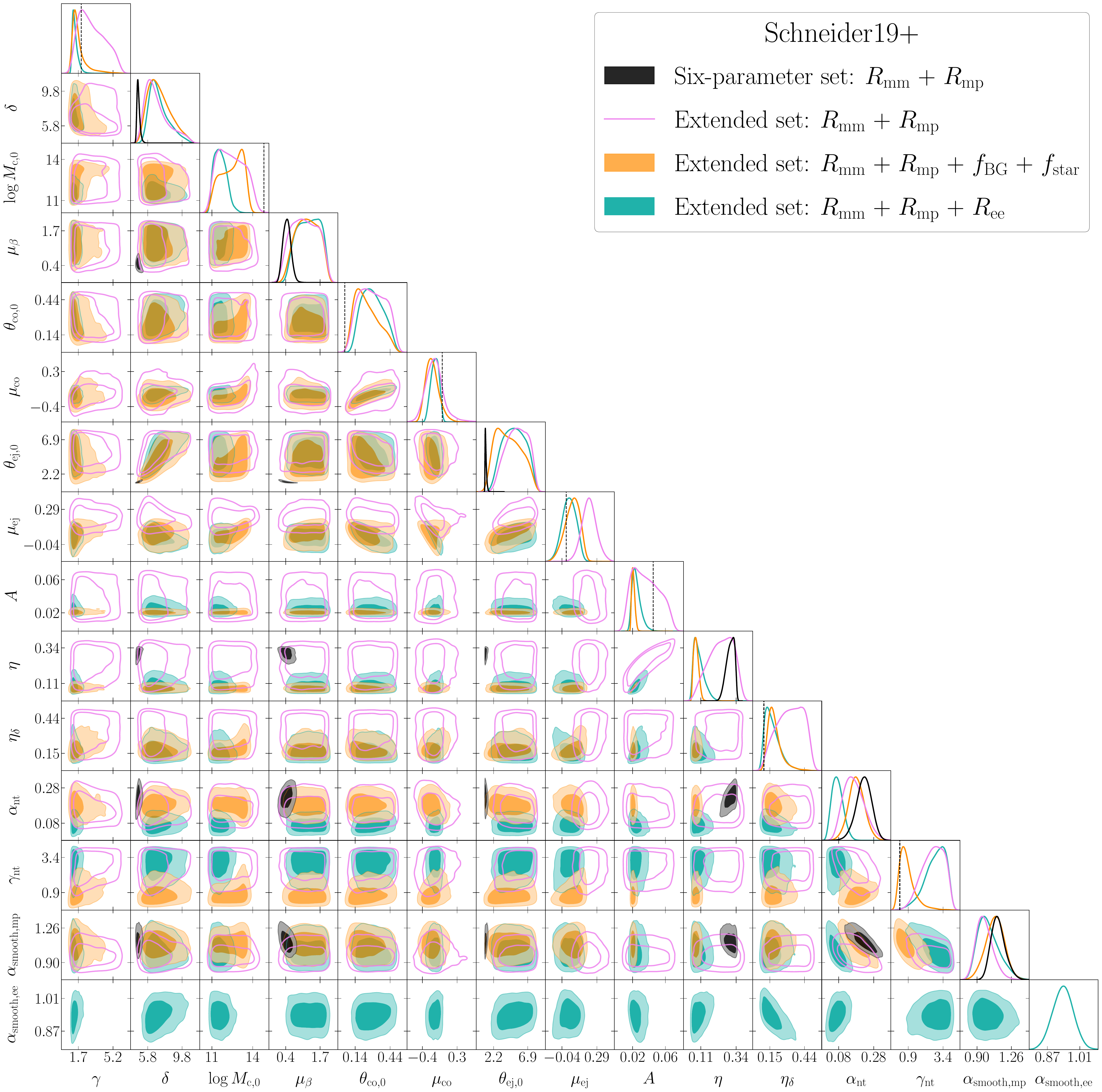}
    \caption{Same as Fig. \protect\refc{fig:post_M20} but for the Schneider19+ model.}
    \label{fig:post_S19}
\end{figure*}

\begin{figure*}
    \centering
    \includegraphics[width=\linewidth]{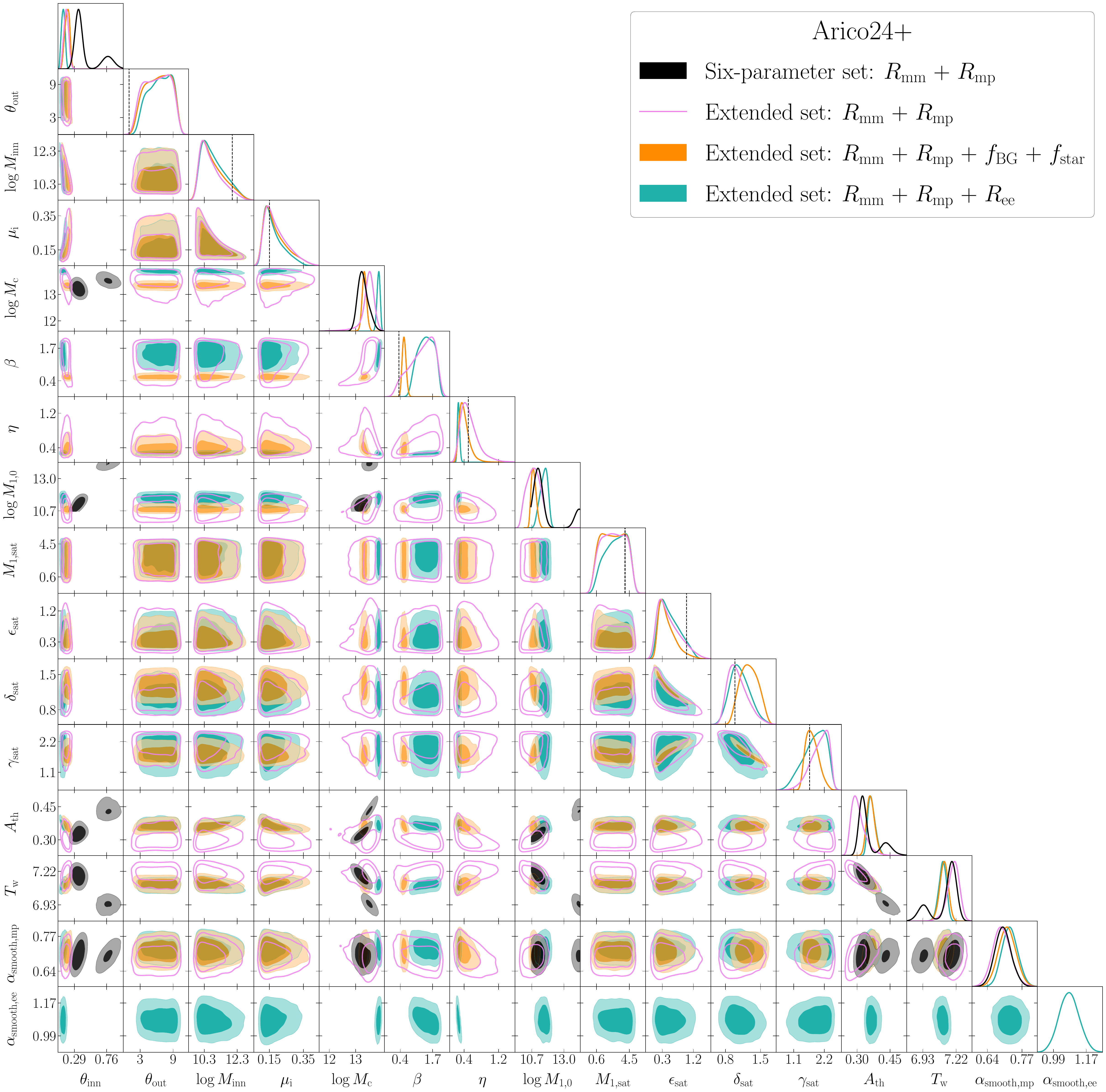}
    \caption{Same as Fig. \protect\refc{fig:post_M20} but for the Arico24+ model.}
        \label{fig:post_A24}
\end{figure*}

In this appendix we present the parameter posteriors for each model,  these are shown in Figs.~\refc{fig:post_M20},~\refc{fig:post_S19}, and~\refc{fig:post_A24}. We also provide a list of best fit parameters in Table~\refc{tab:bf_params}.

\begin{table*}
    \centering
    \begin{tabular}{cllll}
    \toprule
    \textbf{Parameter} & \multicolumn{4}{c}{\textbf{Best fit}}\\
     & \multicolumn{2}{c}{$R_{\rm mm}$ \& $R_{\rm mp}$}&  $R_{\rm mm}$, $R_{\rm mp}\,,f_{\rm BG}\,,f_{\rm star}$& $R_{\rm mm}$,  $R_{\rm mp}$, $R_{\rm ee}$\\
     \cmidrule(lr){2-3} \cmidrule(lr){4-4}  \cmidrule(l){5-5} 
     & Six-parameter set & Extended set & Extended set & Extended set\\\hline
    \multicolumn{5}{c}{\textbf{Mead20+}} \\
    $\epsilon_1$ \dotfill &-0.6& -0.47& -0.47& -0.55\\
    $\epsilon_2$ \dotfill & Fixed to 0 &  -0.27& 0.03&0.52\\
    $\Gamma$ \dotfill & 1.14& 1.12& 1.14& 1.13\\
    $\log (M_0h/\text{M}_\odot)$ \dotfill& 13.45& 12.40& 13.62& 15.56\\
    $\beta$\dotfill & Fixed to 0.6 & 0.24& 0.47& 0.25\\
    $\eta_\text{b}$\dotfill& Fixed to 0.5 & 0.81& 0.55& 0.15\\
    $A_*$\dotfill & Fixed to 0.03& 0.06& 0.08& 0.04\\
    $\log (M_*/\text{M}_\odot)$ \dotfill& Fixed to 12.45& 11.67& 11.51& 14.27\\
    $\alpha$\dotfill& 0.27& 0.27& 0.31& 0.54\\
    $\log (T_\text{w}/$K)\dotfill& 6.60& 6.75& 6.69& 6.58\\
    $\alpha_{\rm smooth,mp}$ \dotfill& 0.70& 0.68& 0.70&0.72\\
     $\alpha_{\rm smooth,ee}$ \dotfill& --& --&--&0.98\\
    
    \hline \multicolumn{5}{c}{\textbf{Schneider19+}}\\
    $\gamma$ \dotfill& Fixed to 2& 3.54& 2.00& 1.14\\
    $\delta$\dotfill& 4.56& 5.63& 4.87& 8.23\\
    $\log (M_\text{c,0}/\text{M}_\odot)$  \dotfill& Fixed 14.83& 13.38&12.71& 11.86\\
    $\mu_\beta$ \dotfill& 0.62& 1.85& 1.43& 1.59\\
    $\theta_\text{co,0}$ \dotfill&Fixed to 0.05& 0.30&0.23& 0.49\\
    $\mu_\text{co}$ \dotfill& Fixed to 0& -0.11& -0.23& -0.07\\
    $\theta_\text{ej,0}$ \dotfill& 1.00& 4.81& 1.08& 5.92\\
    $\mu_\text{ej}$ \dotfill& Fixed to 0& 0.23& -0.02& 0.00\\
    $A$  \dotfill& Fixed to 0.045& 0.03& 0.02& 0.02\\
    $\eta$\dotfill& 0.30& 0.24& 0.07& 0.10\\
    $\eta_\delta$ \dotfill& Fixed to 0.1& 0.25& 0.14& 0.13\\
    $\alpha_\text{nt}$\dotfill&0.24& 0.14& 0.16& 0.07\\
    $\gamma_\text{nt}$ \dotfill& Fixed to 0.3& 2.87& 0.26& 3.91\\
    $\alpha_{\rm smooth,mp}$ \dotfill& 1.06& 0.97& 1.12&0.96\\
    $\alpha_{\rm smooth,ee}$ \dotfill& --& --&--&0.96\\
    
    \hline\multicolumn{5}{c}{\textbf{Arico24+}}\\

    $\theta_\text{inn}$ \dotfill& 0.36& 0.17& 0.15& 0.13\\
     $\theta_\text{out}$ \dotfill& Fixed to 1.0& 9.79& 5.79& 8.89\\
     $\log (M_\text{inn}/\text{M}_\odot)$ \dotfill& Fixed to 12& 10.55& 11.67& 10.98\\
     $\mu_\text{i}$ \dotfill& Fixed to 0.15& 0.17& 0.15& 0.16\\
    $\log (M_\text{c}/\text{M}_\odot)$ \dotfill& 13.10& 13.49& 13.30& 13.79\\
     $\beta$ \dotfill& Fixed to 0.35& 1.96& 0.60& 1.40\\
     $\eta$ \dotfill& Fixed to 0.5 & 0.61& 0.42& 0.26\\
  $\log (M_{1,0}/\text{M}_\odot)$ \dotfill& 11.04& 10.19& 10.51& 10.86\\
    $M_{1,\text{sat}}$ \dotfill& Fixed to 3.98& 4.39& 3.32& 2.40\\
    $\epsilon_\text{sat}$ \dotfill& Fixed to 1.0& 0.59& 0.20& 0.24\\
    $\delta_\text{sat}$ \dotfill& Fixed to 0.99& 0.89& 1.50& 1.38\\
    $\gamma_\text{sat}$\dotfill& Fixed to 1.67& 2.37& 1.47& 1.76\\ 
    $A_{\rm th}$ \dotfill& 0.31& 0.27& 0.36& 0.38\\
    $\log (T_\text{w}/$K)\dotfill& 7.21& 7.23& 7.12& 7.08\\
        $\alpha_{\rm smooth,mp}$ \dotfill& 0.71& 0.66& 0.71&0.70\\
     $\alpha_{\rm smooth,ee}$ \dotfill& --& --&--&1.03\\
    \bottomrule
    \end{tabular}
    \caption{Best fit values for model parameters when fit to measurements from the \textsc{Magneticum} simulations at $z=0.25$. Column (1) shows the model parameters that are varied. Columns (2) and (3) show the best-fit values for the six-parameter and extended set, respectively, when jointly fit to the matter--matter and matter--pressure power spectrum response. Column (4) shows the best-fit parameters when the extended set is used to jointly fit the  matter--matter and matter--pressure power spectrum response along with the mass fractions of bound gas and stars. Column (5) shows the parameters when jointly fitting the matter--matter, matter--pressure, and electron density power spectrum response with the extended set. Model parameters not shown are fixed to fiducial values in Table \protect\refc{tab:model_params}.}
    \label{tab:bf_params}
\end{table*}


\appendix
\label{lastpage}
\end{document}